\documentclass[aps,prx,twocolumn,showpacs]{revtex4-1}
\usepackage{bm}
\usepackage{mathrsfs}
\usepackage{amsmath}
\usepackage{amssymb}
\usepackage{graphicx}
\usepackage{amsfonts}
\usepackage{amsthm}
\usepackage{color}
\usepackage{dcolumn}
\usepackage{txfonts}
\usepackage[caption=false]{subfig}
\begin{document}

\title{Accretion-Driven Squeezing of Fuzzy Dark Matter Halo Cores in the Schr\"odinger
-Newton Framework}

\author{Chon-Fai Kam}
\email{dubussygauss@gmail.com}
\affiliation{Department of Physics, School of Physics, Mathematics and Computing,
The University of Western Australia, Perth 6009, Australia}

\author{Chi-Sheng Chen}
\email{m50816m50816@gmail.com}
\affiliation{
Graduate Institute of
Biomedical Electronics and Bioinformatics,
National Taiwan University
No.1, Sec.4, Roosevelt Road, Taipei, 10617, Taiwan
}

\author{En-Jui Kuo}
\email{kuoenjui@nycu.edu.tw}
\affiliation{
Department of Electrophysics, National Yang Ming Chiao Tung University, Hsinchu, Taiwan, R.O.C.}

\begin{abstract}
We investigate the impact of accretion onto supermassive black holes on the density profiles of a fuzzy dark matter soliton core at the center of a dark matter halo. Treating the supermassive black hole at the center of a galaxy as a point mass, we numerically solve the Schr\"odinger-Newton equation for the scalar field. We find that the time-dependent perturbation has a significant squeezing effect on the soliton density profile, which both reduces the size of the core and increases the central density. This finding provides insights into how black hole growth influences fuzzy dark matter structures, potentially addressing discrepancies in galactic core observations.
\end{abstract}

\maketitle

\section{Introduction}
Despite compelling circumstantial evidence for its existence, the precise particle nature of dark matter---which constitutes approximately 27\% of the Universe's energy density---remains unknown and continues to be one of the most profound mysteries in physics \cite{bertone2018new}. Dark matter is about five times more abundant than baryonic matter, such as protons and neutrons, which together account for less than 5\% of the Universe's energy budget. This disparity allows dark matter to dominate the formation of large-scale cosmic structures \cite{frenk2012dark} and to govern the evolution of galaxies within merging dark matter halos \cite{vogelsberger2020cosmological, de2020dark}. Although its fundamental nature is still elusive, a wide range of observational evidence strongly indicates that dark matter cannot be composed of any known particles within the Standard Model of particle physics \cite{bertone2018new}. This has led to the proposal of various theoretical candidates, including sterile neutrinos, ultralight bosons, and even modifications to gravity. Among these, the cold dark matter (CDM) paradigm---particularly models involving weakly interacting massive particles (WIMPs)---has emerged as one of the most promising frameworks, as it successfully reproduces the large-scale structure of the Universe in cosmological simulations \cite{davis1985evolution, springel2005simulations}.

The WIMP model has long dominated dark matter research, prized for its theoretical elegance and its alignment with cosmological observations. Yet, the absence of direct experimental evidence has increasingly undermined its credibility. Despite extensive efforts using highly sensitive underground detectors such as LUX, XENON, and CDMS, no definitive WIMP signals have been detected. Furthermore, the Large Hadron Collider has failed to uncover supersymmetric particles often associated with WIMPs, while the ``neutrino floor''---a fundamental limit where neutrino signals overwhelm potential WIMP detections---threatens the viability of future searches. These challenges highlight the need to explore alternatives.

Even the CDM model itself, despite its success in describing large-scale cosmic structure, encounters significant challenges on small scales ($\lesssim$ hundreds of kpc) \cite{weinberg2015cold, bullock2017small}. One major issue is the ``cusp-core'' (CC) problem \cite{moore1994evidence}, where collisionless $N$-body simulations predict a steep central density profile for CDM halos, typically $\rho\propto r^{-x}$ with $1\lesssim x\lesssim 1.5$. This prediction contradicts observations of dwarf galaxies, irregulars, and low surface brightness galaxies, which exhibit flat, core-like profiles consistent with $\rho\propto r^0$ \cite{navarro1996structure, navarro2010diversity, saburova2014surface}. Another challenge is the ``missing satellite'' (MSP) problem, referring to the discrepancy between the number of sub-halos predicted by simulations and the observed number of satellite galaxies. CDM simulations suggest the Local Group should host around 1,000 satellites, yet only about 50 dwarf galaxies have been observed \cite{moore1999dark, klypin1999missing}. These small-scale tensions, combined with the lack of direct detection of CDM particles such as WIMPs with masses in the GeV to hundreds of GeV range \cite{agnes2018low}, underscore the need to explore alternative dark matter models.

These null results have revitalized interest in wave-like dark matter candidates, such as ultralight bosons. Ultralight bosons such as axions have emerged as compelling alternatives. With masses as low as $10^{-22}$ eV and no electromagnetic interaction, they are nearly undetectable by conventional means yet could account for up to 85\% of the Universe's dark matter. Unlike CDM, their wave-like nature suppresses small-scale structure formation, offering elegant solutions to cosmological puzzles such as the overly dense centers of galaxies. These particles also open new avenues in black hole physics. Through superradiant instability, ultralight bosons can extract angular momentum from rotating black holes, forming boson clouds that emit continuous, nearly monochromatic gravitational waves. Such emissions provide novel observational signatures, and the presence of these bosons could constrain black hole spin rates---offering indirect evidence for their existence.

From a detection standpoint, ultralight bosons invite a shift beyond traditional particle physics. Gravitational wave observatories like LIGO, Virgo, and KAGRA may detect signals from boson clouds, while multi-messenger astronomy could reveal their influence on binary star systems or trigger exotic phenomena such as ``boson novas.'' Theoretical appeal is strong as well: these particles naturally arise in extensions of the Standard Model, including supersymmetry and string theory, and their quantum wave behavior on cosmic scales hints at profound connections between particle physics and cosmology.

One particularly intriguing model is fuzzy cold dark matter (FCDM), which posits that galactic dark matter consists of ultralight scalar bosons with negligible self-interactions and masses on the order of $10^{-22}$ eV \cite{hu2000fuzzy, goodman2000repulsive}. In this scenario, the bosons form a Bose-Einstein condensate, sharing a single wave function. Their large Compton wavelengths---on the order of light-years---induce wave-like behavior at galactic scales, smoothing out central cusps in halos due to quantum pressure and the Heisenberg uncertainty principle \cite{mocz2018schrodinger, mocz2019first}. As a result, FCDM halos exhibit soliton cores of a few kiloparsecs, offering a natural resolution to the cusp-core problem. However, the model faces constraints: boson masses below $10^{-24}$ eV would erase large-scale structure, while masses above $10^{-20}$ eV would recover classical CDM behavior, making it difficult to tightly constrain the particle mass \cite{mocz2018schrodinger}. Recent observations from the Lyman-$\alpha$ forest and James Webb Space Telescope \cite{maio2023jwst} further refine these bounds, supporting FCDM's viability for resolving small-scale tensions---e.g., by predicting core densities $\rho_c \sim 10^7-10^9 M_\odot/\mathrm{kpc}^3$ consistent with dwarf galaxy data.

Observational evidence from instruments such as the Event Horizon Telescope, Chandra X-ray Observatory, and James Webb Space Telescope confirms that supermassive black holes reside at the centers of nearly all large galaxies, including the Milky Way, with masses ranging from millions to billions of solar masses \cite{ferrarese2000fundamental, gebhardt2000relationship}. Understanding how dark matter halos evolve under the gravitational influence of these central black holes is essential for exploring the connection between galaxies and their dark matter environments \cite{reddick2013connection}. While fully relativistic treatments of scalar field dark matter near supermassive black holes are possible \cite{barranco2012schwarzschild, barranco2017self, avilez2018possibility, hui2019black}, this work focuses on the halo's mass density profile at distances far from the Schwarzschild radius, where the black hole can be approximated as a point mass \cite{davies2020fuzzy}. Given that most galactic cores feature accreting black holes with significant luminosity, investigating the impact of black hole accretion on dark matter halo structure is particularly relevant---and forms the central aim of the present study. In this paper, we numerically solve the Schr\"odinger-Newton equations to model how time-dependent accretion perturbs FCDM soliton cores, revealing a significant squeezing effect that reduces core sizes and boosts central densities---potentially exacerbating or resolving cusp-core discrepancies in observed galaxies. Section II introduces the model, followed by steady-state solutions, numerical methods, and restrictions on FDM parameters.

\section{The Model}
To model fuzzy dark matter (FDM) in galactic halos, we treat the ultralight bosons as a classical scalar field $\Psi(\mathbf{r},t)$ due to their high occupation numbers ($\gg 1$), which justifies a mean-field approximation. This field obeys the Schr\"odinger-Newton (SN) equation, coupling quantum evolution to self-gravitation via Poisson's equation---ideal for capturing wave-like interference on kpc scales while incorporating Newtonian gravity. To describe the dark matter halo density profiles around an accreting supermassive black hole, we consider a scalar field dark matter model in which dark matter consists of ultralight fuzzy dark matter (FDM) particles. Given the high occupation numbers in galactic halos, FDM can be treated as a scalar field $\Psi(\mathbf{r},t)$ minimally coupled to gravity, obeying the Schr\"{o}dinger-Newton equation \cite{tod1999analytical}
\begin{equation}
    i\hbar\partial_t\Psi(\mathbf{r},t) = \left(-\frac{\hbar^2\nabla^2}{2m}+m\Phi(\mathbf{r},t)+mV(r,t)\right)\Psi(\mathbf{r},t),
\end{equation}
where $m\approx 10^{-22}$eV is the FDM boson mass, $\Phi(\mathbf{r},t)$ is the self-potential of the wave function, which obeys Poisson's equation for the Newtonian gravitational field
\begin{equation}
    \nabla^2\Phi(\mathbf{r},t)=4\pi G |\Psi(\mathbf{r},t)|^2,
\end{equation}
and $V(r,t)$ is a time-dependent black hole potential given by
\begin{equation}
V(r,t)=-\frac{GM(t)}{r}.
\end{equation}
Here, $M(t)$ is the time-dependent black hole mass, and $G$ is the gravitational constant. This non-relativistic setup is valid at distances $r \gg r_s = 2GM_0/c^2$ from the black hole, where relativistic effects are negligible, allowing us to focus on halo-scale perturbations ($\sim$ kpc). To model the central accreting SMBH as a time-dependent perturbation, we adopt the standard Shakura-Sunyaev disk (SSD) model \cite{shakura1973black, smole2015smbh}, in which the black hole mass varies as
\begin{equation}
    M(t)\equiv M_0\exp\left(\tilde{\lambda}\frac{1-\epsilon}{\epsilon}\frac{t}{t_E}\right)\equiv M_0e^{\alpha t},
\end{equation}
where $\tilde{\lambda} \equiv L/L_E$ is the \textit{Eddington ratio}, which relates the luminosity $L$ of an accreting super-massive black hole with the \textit{Eddington luminosity} $L_E\equiv 4\pi cGMm_p/\sigma_T$, with $\sigma_T$ being the Thomson scattering cross-section for the electron, and $m_p$ being the mass of the proton. Here, $\epsilon\approx 0.1$ is the radiative efficiency of accretion, and $t_E\approx 4.5\times 10^7$yrs is the Salpeter time scale --- the e-folding time-scale for super-massive black hole growth. As the Poisson equation for the self-potential $\Phi(\mathbf{r},t)$ is solved by the integral
\begin{equation}
    \Phi(\mathbf{r},t)=-G\int\frac{|\Psi(\mathbf{r}^\prime,t)|^2}{|\mathbf{r}-\mathbf{r}^\prime|}d^3r^\prime,
\end{equation}
the scalar field $\Psi(\mathbf{r},t)$ obeys the following differential-integral equation
\begin{align}
    i\hbar\partial_t\Psi(\mathbf{r},t) &= \left(-\frac{\hbar^2\nabla^2}{2m}-Gm\int\frac{|\Psi(\mathbf{r}^\prime,t)|^2}{|\mathbf{r}-\mathbf{r}^\prime|}d^3r^\prime\right.\nonumber\\
    &\left.-Gm\frac{M(t)}{r}\right)\Psi(\mathbf{r},t).
\end{align}
We assume spherical symmetry around the black hole, justified for near-equilibrium solitons with azimuthal symmetry, reducing the system to a 1+1D form. For spherically symmetric solutions, one obtains a 1+1 dimensional differential integral equation which governs the spherically symmetric solutions of the scalar field
\begin{align}\label{1plus1}
    i\hbar\partial_t\Psi(r,t) &= -\frac{\hbar^2}{2mr}\frac{\partial^2}{\partial r^2}(r\Psi(r,t))-Gm\left(\frac{4\pi}{r}\int_0^r|\Psi(r^\prime,t)|^2r^{\prime 2}dr^\prime\right. \nonumber\\
    &\left.+ 4\pi\int_r^\infty|\Psi(r^\prime,t)|^2r^\prime dr^\prime+\frac{M_0e^{\alpha t}}{r}\right)\Psi(r,t)\nonumber\\
    &\equiv\hat{H}(r,t)\Psi(r,t).
\end{align}
This framework enables numerical studies of how accretion-driven growth perturbs FDM soliton cores, as explored in subsequent sections. We first discuss the steady solutions in the following sections.

\section{The steady-state spherical solution}
The steady-state spherical solution $\phi(r)$ can be solved by the ansatz $\Psi(r,t)=e^{-iEt/\hbar}\phi(r)$, so that the Schrodinger-Poisson equation for a time-independent black hole potential becomes
\begin{subequations}
\begin{align}
    E\phi(r)&=\left(-\frac{\hbar^2}{2m}\nabla^2+m\Phi(r)+mV(r)\right)\phi(r),\\
    \nabla^2\Phi(r)&=4\pi G|\phi(r)|^2.
\end{align}
\end{subequations}
Or equivalently
\begin{figure}[tbp]
\begin{center}
\subfloat[$\lambda=0$\label{sfig:Fig1a}]{%
  \includegraphics[width=0.48\columnwidth]{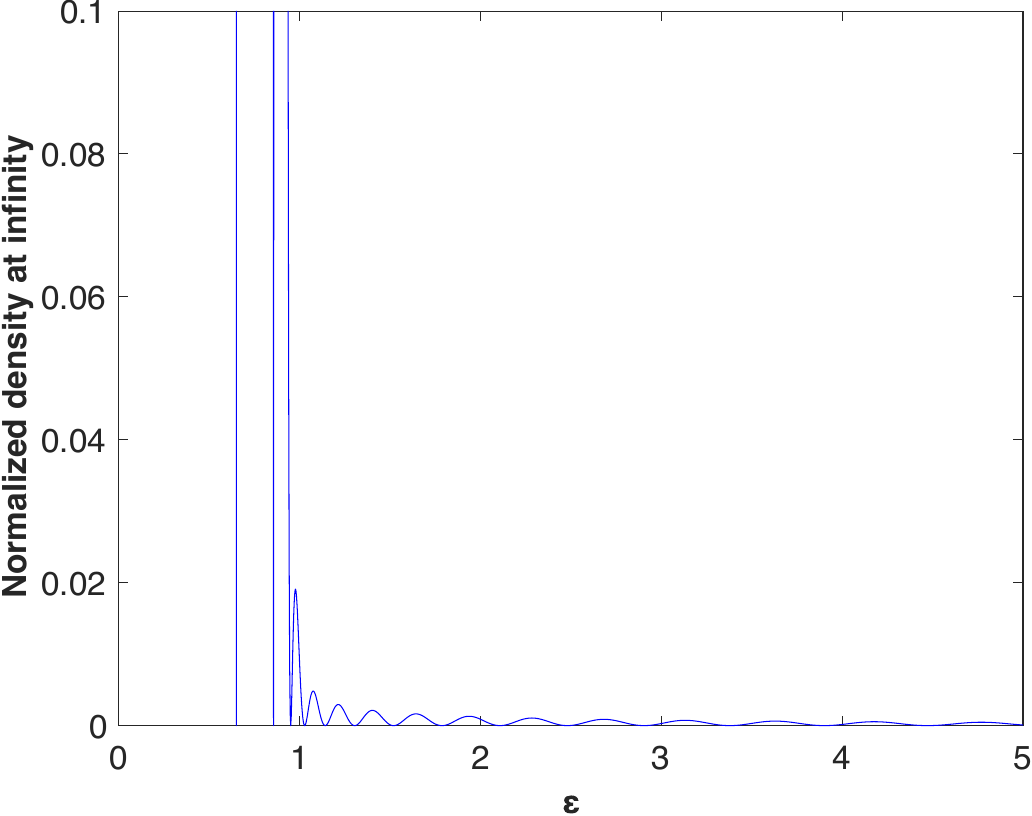}%
}\hfill
\subfloat[$\lambda=0.1$\label{sfig:Fig1b}]{%
  \includegraphics[width=0.48\columnwidth]{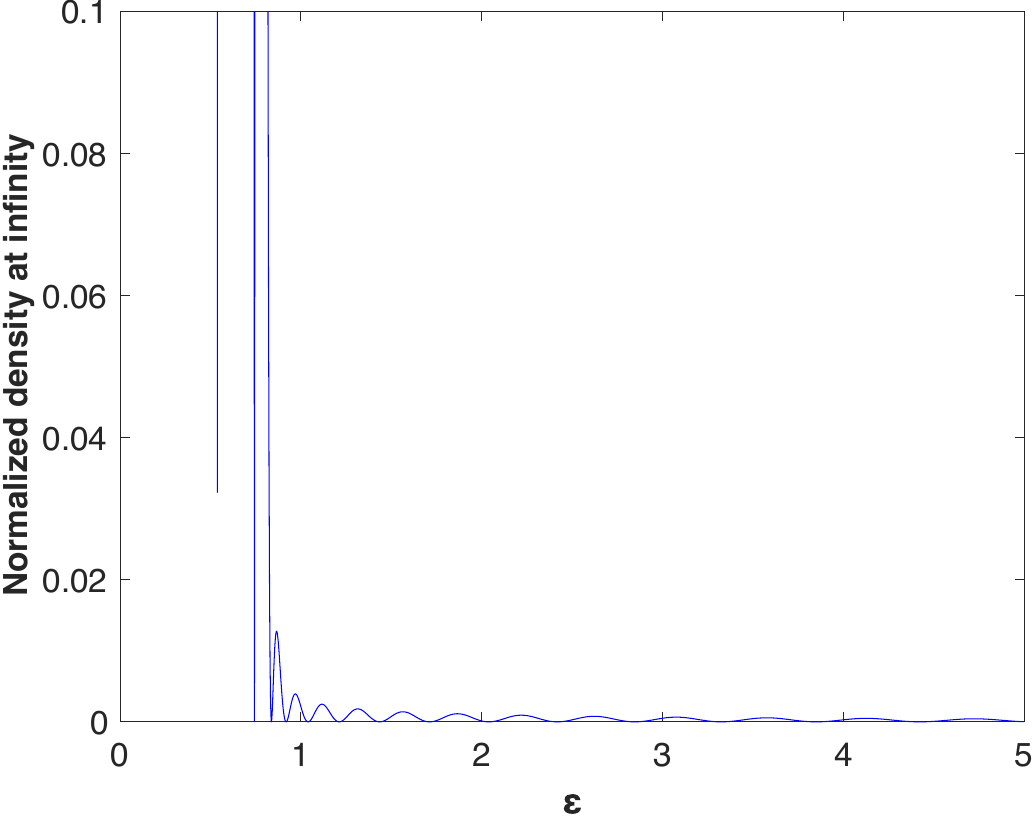}%
}\hfill
\subfloat[$\lambda=0.5$\label{sfig:Fig1c}]{%
  \includegraphics[width=0.48\columnwidth]{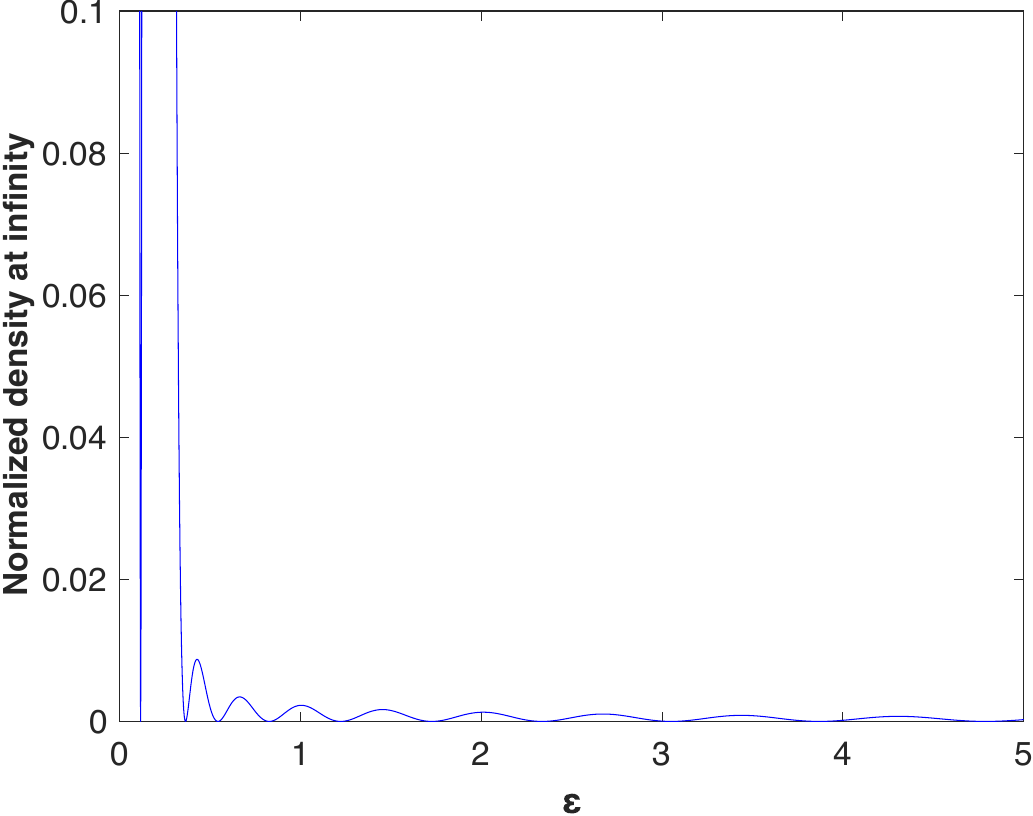}%
}\hfill
\subfloat[$\lambda=1.0$\label{sfig:Fig1d}]{%
  \includegraphics[width=0.48\columnwidth]{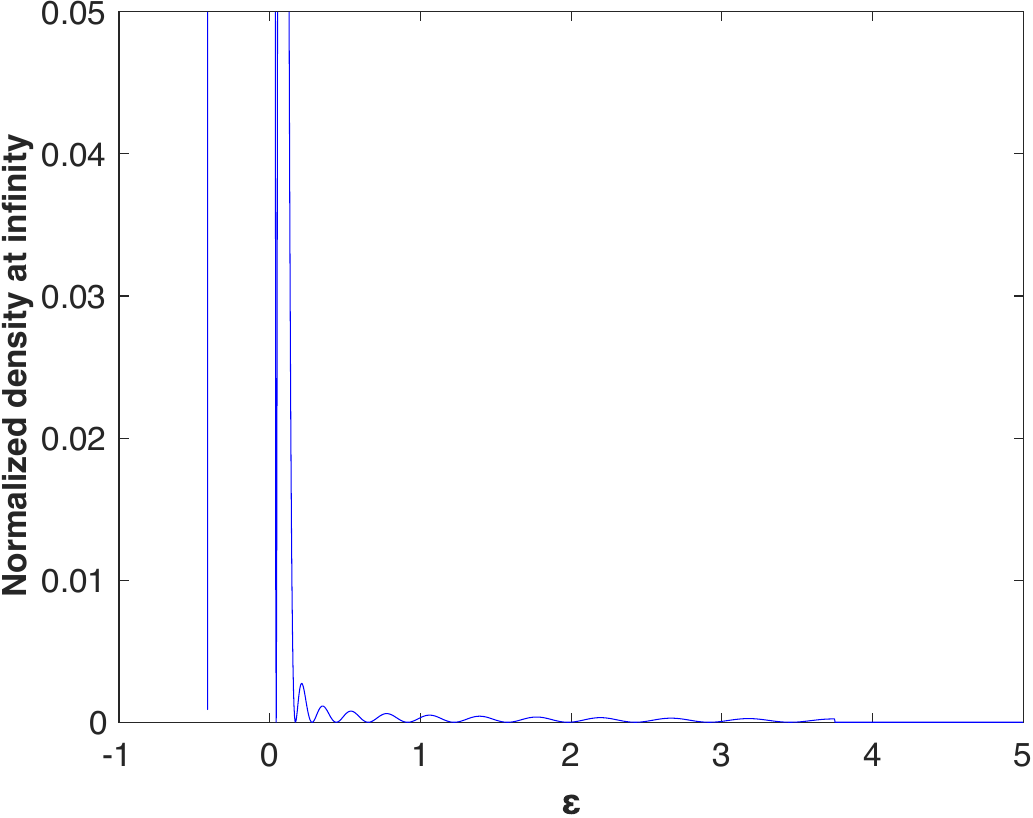}%
}
\caption{Schematic of the asymptotic normalized densities $\tilde{\phi}(u)^2$ for $u\rightarrow\infty$ as functions of the parameter $\epsilon$ for different $\lambda$ values, where $\lambda\equiv GM_0m/(\hbar c)$ is the dimensionless mass of the black hole. Here, we fixed the dimensionless distance $u$ to $15$ when evaluating the asymptotic normalized densities $\tilde{\phi}(u)^2$.}
\label{fig:Fig1}
\end{center}
\end{figure}\begin{subequations}
\begin{align}
    \frac{\hbar^2}{2m}\frac{d^2}{d r^2}\left(r\phi(r)\right)&=\left(rm\Phi(r)-rE-GmM_0\right)\phi(r),\\
    \frac{d^2}{dr^2}\left(r\Phi(r)\right)&=4\pi Gr\phi(r)^2,
\end{align}
\end{subequations}
which are subjected to the initial conditions $\phi(0)=1$, $\phi^\prime(0)=\Phi(0)=\Phi^\prime(0)=0$ and the boundary condition $\phi(\infty)=0$. We use the shooting method to solve the above set of differential equations, which starts by a trial solution at $r=0$, and tries to reach the asymptotic boundary condition at $r=\infty$. To simplify numerical computation, one may introduce a set of dimension-less variables via
\begin{equation}
    u\equiv \frac{r}{L_0}, \Tilde{\Phi}\equiv \frac{\Phi}{c^2},\epsilon\equiv \frac{E}{mc^2},
\end{equation}
so that the renormalized time-independent coupled SP equations becomes
\begin{figure}[tbp]
\begin{center}
\subfloat[$\lambda=0$\label{sfig:Fig2a}]{%
  \includegraphics[width=0.49\columnwidth]{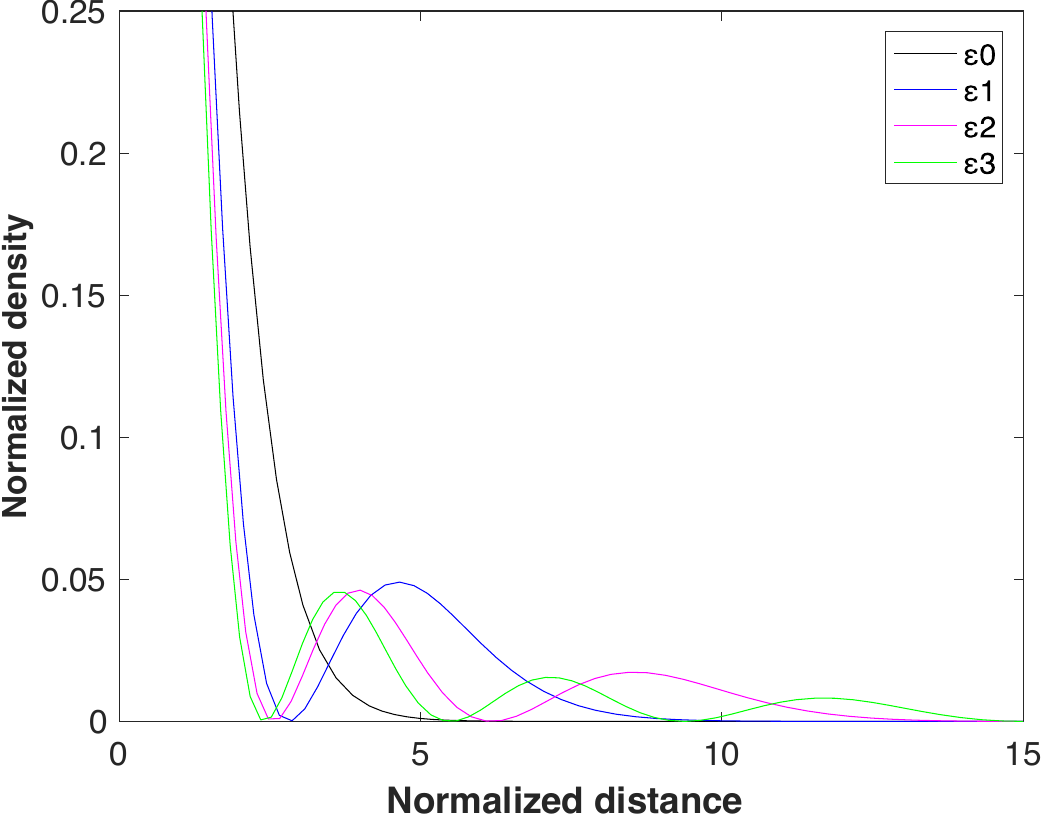}%
}\hfill
\subfloat[$\lambda=0.1$\label{sfig:Fig2b}]{%
  \includegraphics[width=0.49\columnwidth]{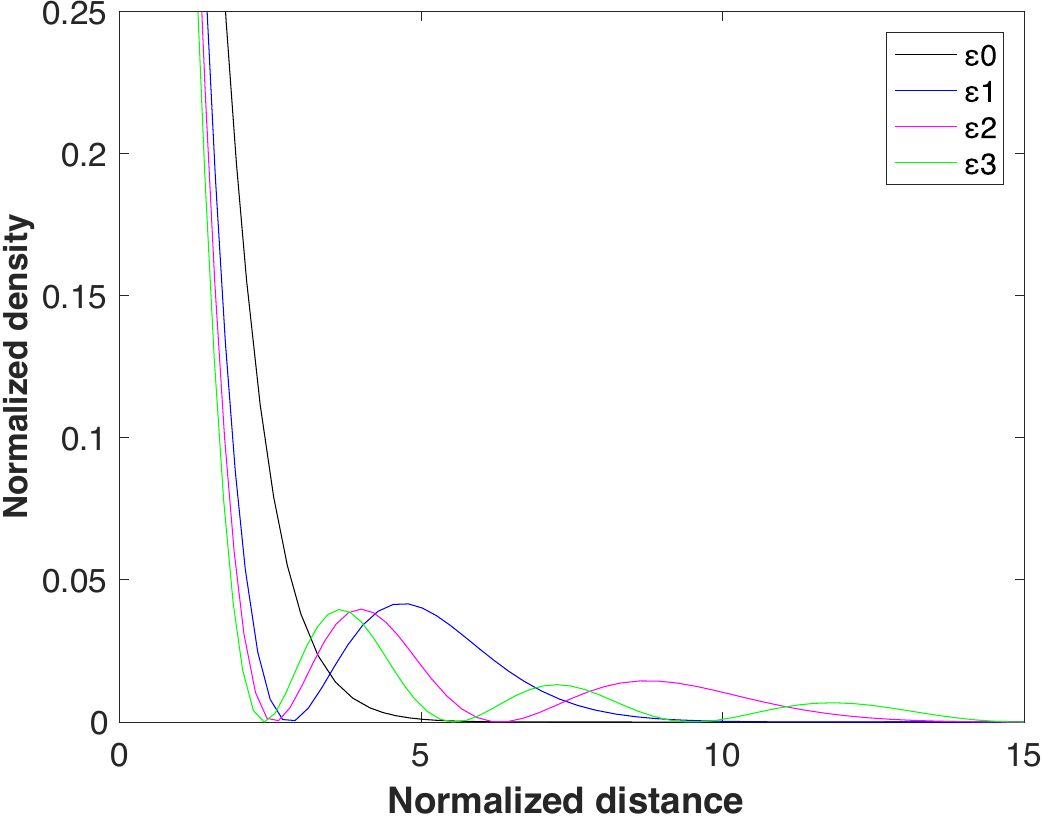}%
}\hfill
\subfloat[$\lambda=0.5$\label{sfig:Fig2c}]{%
  \includegraphics[width=0.49\columnwidth]{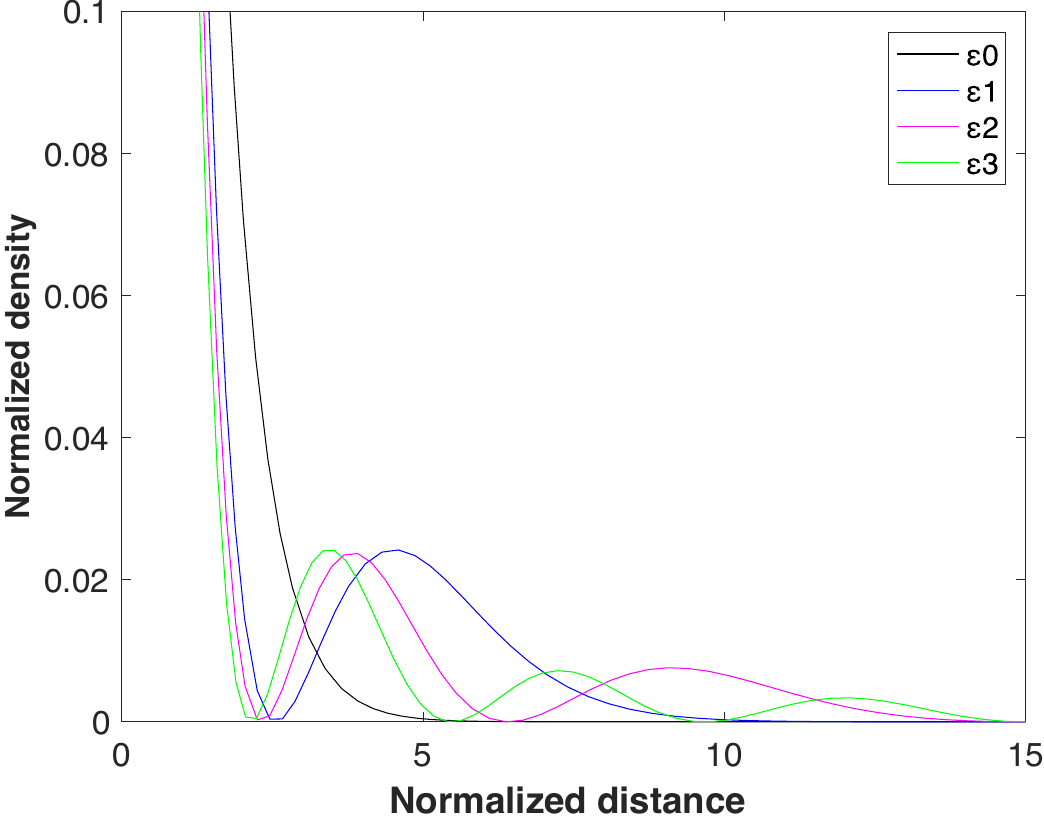}%
}\hfill
\subfloat[$\lambda=1.0$\label{sfig:Fig2d}]{%
  \includegraphics[width=0.49\columnwidth]{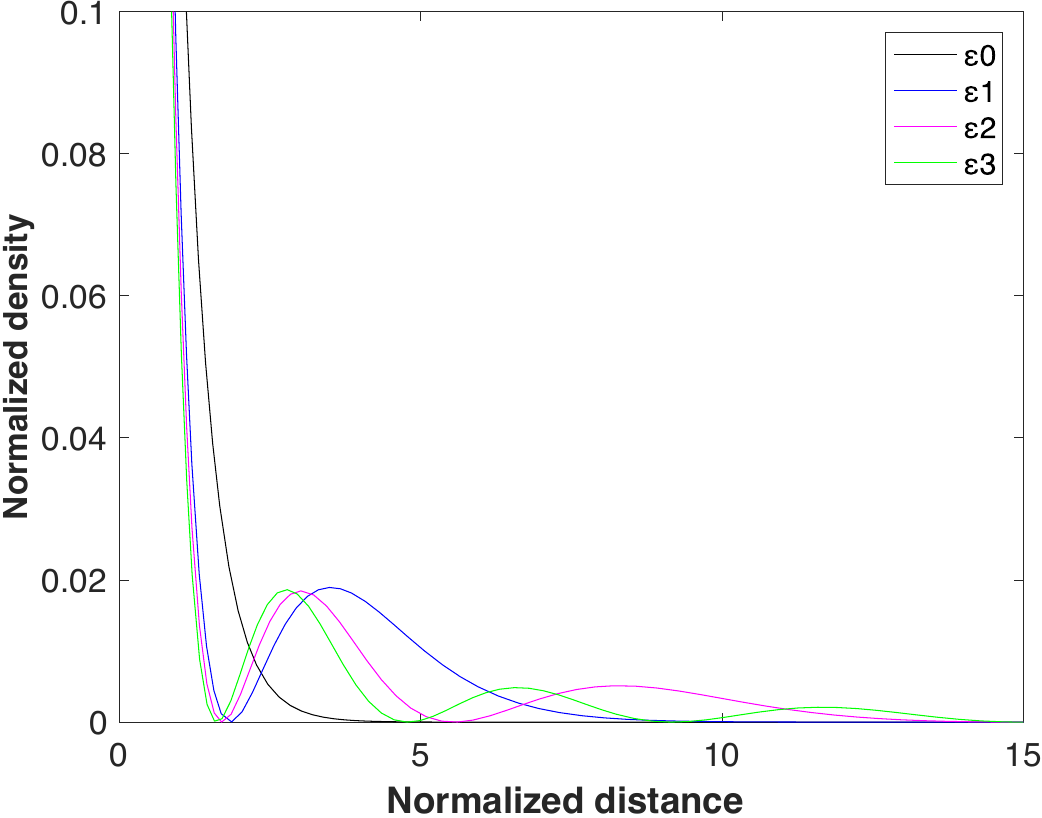}%
}
\caption{Schematic of the normalized densities $\tilde{\phi}^2_n\equiv 4\pi G\hbar^2\phi_n^2/(m^2c^4)$ as functions of the normalized distance $u\equiv mcr/\hbar$ for different $\lambda$ values, where $n$ is an integer which labels the discrete eigenvalues $\epsilon_n$, and $\lambda\equiv GM_0m/(\hbar c)$ is the dimensionless mass of the black hole. Here, we only plot the normalized densities for the first four eigenvalues.}
\label{fig:Fig2}
\end{center}
\end{figure}\begin{subequations}
\begin{align}
    \frac{d^2}{du^2}(u\tilde{\phi})&=2u\left(\tilde{\Phi}-\epsilon-\frac{\lambda}{u}\right)\tilde{\phi},\\
    \frac{d^2}{du^2}(u\tilde{\Phi})&=u\tilde{\phi}^2,
\end{align}
\end{subequations}
where $L_0\equiv \hbar/(mc)$, $\lambda\equiv GM_0m/(\hbar c)$, and $\tilde{\phi}\equiv \sqrt{4\pi G}L_0\phi/c$. As an example, for $m\approx 10^{-22}$eV, one obtains $L_0\approx 0.1380$\:m, $\lambda\approx 5.3832\times 10^{-27}M_0$, and $\tilde{\phi}\approx 1.3326\times 10^{-14}\phi$. The above coupled SP equations can be written as a system of first-order ordinary differential equation as
 \begin{table}[tbp]
 \centering
 \begin{tabular}{ |c|c|c|c|c| }
 \hline
  $\lambda$ & $\epsilon_0$ & $\epsilon_1$ & $\epsilon_2$ & $\epsilon_3$\\ 
 \hline
  0 & 0.649599900660242 & 0.855382257832184 & 0.950376 & 1.027724 \\ 
  \hline
  0.1 & 0.537515694614713 & 0.742953148283380 & 0.836921 & 0.918315\\ 
 \hline
 0.5 & 0.116385322639802 & 0.360002326932728 & 0.459582 & 0.560518 \\ 
 \hline
 1.0 & -0.411778217857106 & 0.043028169337089 & 0.171107 & 0.281542\\ 
 \hline
\end{tabular}
\caption{Table of the four smallest eigenvalues $\epsilon_n$ which obey the asymptotic boundary condition $\tilde{\phi}(\infty)=0$ for different values of $\lambda$.}
\label{tab:table1}
\end{table}\begin{subequations}

\begin{figure}[tbp]
\begin{center}
\includegraphics[width=0.5\textwidth]{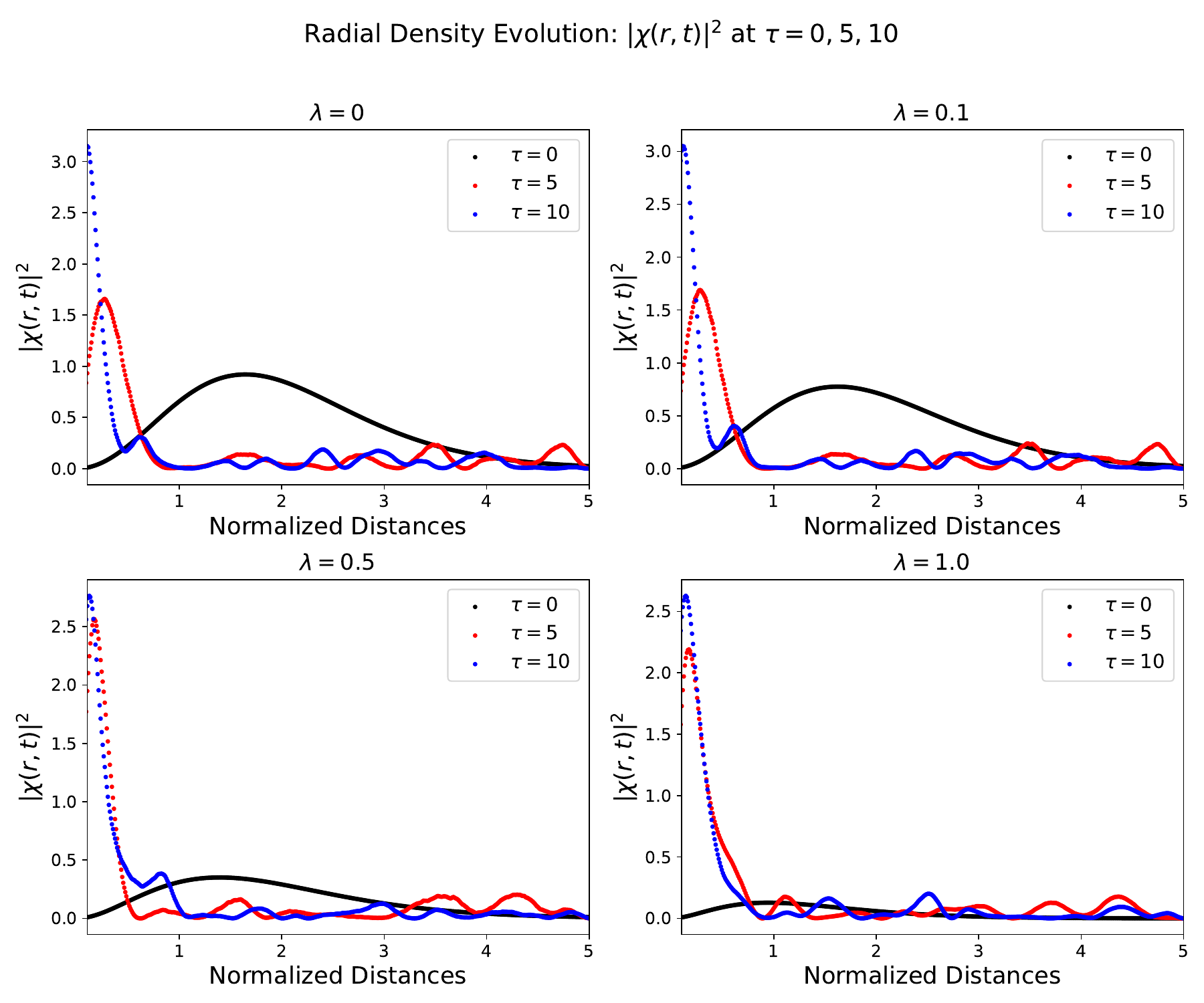}
\caption{Radical densities $|\tilde{\chi}(u,\tau)|^2 \equiv 4\pi G\hbar^2/(m^2c^4)|\chi(r,t)|^2$ as functions of radial distance and time in dimensionless units, for an initial 0-node (ground state) solution. We present results for $\lambda \equiv GM_0m/(\hbar c) = 0$, $0.1$, $0.5$, and $1.0$, with fixed $\alpha = 0.1$ in all cases. It is reasonable to observe that as the evolution begins, the mass collapses toward the origin and the tails diminish, since $|\chi(r,t)|^2$ must remain a conserved quantity. We choose $dt = 5 \times 10^{-4}$ for a total of $2 \times 10^4$ steps. In App.\:\ref{sec:con}, we provide a convergence test to demonstrate the accuracy of this simulation.}
\label{fig:Fig3}
\end{center}
\end{figure}

\begin{align}
    y_1^\prime &=y_2,\\
    y_2^{\prime}&=-\frac{2}{u}y_2+2\left(y_3-\epsilon-\frac{\lambda}{u}\right)y_1,\\
    y_3^\prime&=y_4,\\
    y_4^\prime&=-\frac{2}{u}y_4+y_1^2,
\end{align}
\end{subequations}
where $y_1\equiv\tilde{\phi}$, $y_3\equiv \tilde{\Phi}$. One may also simplify the the coupled SP equations by introducing $y_1\equiv u\tilde{\phi}$ and $y_3\equiv u\tilde{\Phi}-\lambda$, which obey the following set of coupled differential equations
\begin{subequations}
\begin{align}\label{NewSP1}
    y_1^{\prime\prime}&=2\left(\frac{y_3}{u}-\epsilon\right)y_1,\\
    y_3^{\prime\prime}&=\frac{1}{u}y_1^2,\label{NewSP2}
\end{align}
\end{subequations}
where the initial conditions are $y_1(0)=0$, $y_1^\prime(0)=1$, $y_3(0)=-\lambda$, and $y_3^\prime(0)=0$, and the asymptotic boundary condition at infinity is $\lim_{u\rightarrow\infty}u^{-1}y_1=0$. Eqs.\:\eqref{NewSP1} - \eqref{NewSP2} can be written as a system of first-order ordinary differential equations as

\begin{subequations}
\begin{align}
    y_1^\prime&=y_2,\\
    y_2^\prime&=2\left(\frac{y_3}{u}-\epsilon\right)y_1,\\
    y_3^\prime&= y_4,\\
    y_4^\prime&= \frac{1}{u}y_1^2.
\end{align}
\end{subequations}
where the initial conditions are $y_1(0)=0$, $y_2(0)=1$, $y_3(0)=-\lambda$, and $y_4(0)=0$.

Table~\ref{tab:table1} enumerates the four lowest discrete eigenvalues $\epsilon_n$ ($n=0,1,2,3$) of the time-independent nonlinear Schr\"{o}dinger-Poisson system (Eqs.~(8)--(9)), obtained via the shooting method for varying dimensionless black hole masses $\lambda \equiv G M_0 m / (\hbar c)$. These eigenvalues characterize the bound-state spectrum of the equilibrium soliton profiles $\tilde{\phi}_n(u)$, indexed by the number of radial nodes and satisfying the asymptotic boundary condition $\tilde{\phi}(\infty) = 0$. In contrast to the linear Schrr\"{o}dinger equation, where exact analytic eigenvalues enable precise exponential decay at infinity, the nonlinear coupling between the scalar field and its self-gravitational potential via Poisson's equation precludes closed-form solutions. Consequently, the boundary condition cannot be satisfied with infinite precision, as it demands arbitrarily accurate determination of the nonlinear eigenvalues; small deviations ($\delta \epsilon \sim 10^{-16}$) can induce unphysical oscillations or incomplete decay in the far-field tails. To mitigate this, our computations retain 16 significant digits throughout the shooting iterations, achieving relative errors below $10^{-6}$ for normalized distances $u \leq 15$ (as verified in Fig.~1). This precision ensures robust initial conditions for the time-dependent evolutions, where ground-state ($n=0$) profiles exhibit stable, node-free central densities ($\rho \propto r^0$) consistent with fuzzy dark matter phenomenology.

From Figs.~\ref{fig:Fig1} and \ref{fig:Fig2}, it is evident that, for a fixed dimensionless black hole mass $\lambda \equiv GM_0 m / (\hbar c)$, the time-independent coupled Schr\"odinger-Poisson equations admit a discrete spectrum of eigenvalues $\epsilon_n$ ($n = 0, 1, 2, \ldots$) that satisfy the asymptotic boundary condition $\tilde{\phi}(\infty) = 0$. These eigenvalues, along with their corresponding eigenfunctions $\tilde{\phi}_n(u)$, are indexed by the number of radial nodes in $\tilde{\phi}_n(u)$, where $u \equiv m c r / \hbar$ denotes the normalized radial distance.

Fig.~\ref{fig:Fig1} illustrates the asymptotic behavior of the normalized densities $\tilde{\phi}^2(u)$ as $u \to \infty$, revealing a unique set of admissible $\epsilon_n$ for each $\lambda$, determined via the shooting method to enforce exponential decay at large $u$. Meanwhile, Fig.~\ref{fig:Fig2} displays the associated normalized density profiles $\tilde{\phi}_n^2(u) \equiv 4\pi G \hbar^2 \phi_n^2 / (m^2 c^4)$ for the lowest four eigenstates ($n=0,1,2,3$) across varying $\lambda$, highlighting the progressive compression and steepening of the profiles with increasing $\lambda$. Notably, the ground state ($n=0$), characterized by a node-free profile and a flat central density $\rho \propto r^0$, is linearly stable against perturbations and corresponds to the equilibrium soliton core in fuzzy dark matter theory~\cite{mocz2018schrodinger,mocz2019first}. In this framework, the dimensionless black hole mass $\lambda$ thus governs the structure of the soliton density profile, with higher $\lambda$ inducing a reduction in core size and an enhancement of central density, as quantified by the eigenvalues listed in Table~\ref{tab:table1}. This baseline equilibrium configuration serves as the initial condition for the subsequent time-dependent evolutions under accretion-driven perturbations.

While this steady-state solution assumes a static central point mass and spherical symmetry—reasonable for equilibrated FDM halos at kpc scales prior to significant accretion—it provides a robust baseline for perturbation studies. 
Spherical symmetry is appropriate for near-equilibrium FDM solitons in isolated halos, where azimuthal averaging suppresses asymmetries, though real galaxies may introduce mild deviations due to rotation or mergers~\cite{reddick2013connection}. 
The point-mass approximation for the black hole holds at halo scales, $r \sim$ kpc $\gg r_s \sim 10^{-6}$ pc for $M_0 \sim 10^6$ - $10^9$ M$_\odot$, neglecting relativistic effects near the event horizon.

An infinite family of normalizable spherically symmetric solutions exists, but linear stability analyses show that only the ground state ($n=0$) is stable against perturbations, while excited states decay to it~\cite{mocz2018schrodinger}. 
This justifies our focus on the $n=0$ profile as the initial condition for time-dependent evolutions in Sec.\:IV, as higher modes would not persist in realistic halos.

Numerical convergence of the shooting method was verified by ensuring asymptotic decay $\phi(\infty)=0$ within $10^{-6}$ relative error for $u$ up to 15 (Fig.~\ref{fig:Fig1}). 
Physically, these ground-state profiles exhibit flat central densities ($\rho \propto r^0$), consistent with FDM resolutions to the cusp-core problem and observational core sizes of $\sim$1 kpc in dwarf galaxies~\cite{schive2014cosmic}. 
However, the absence of self-interactions (valid for $m \sim 10^{-22}$ eV) assumes negligible quartic terms, which could stabilize excited states if included.

While these steady-state solutions provide valuable insights into the equilibrium configurations of FDM halos under a static central mass, they represent only the initial state before accretion begins. To capture the time-dependent effects of black hole growth—such as the squeezing of the soliton core—we must evolve the system dynamically using numerical methods. In the following section, we outline the computational approaches employed to solve the time-dependent Schrödinger-Newton equations, building directly on the ground-state profiles derived here as initial conditions.

\section{The Numerical Method}

Building on the steady-state solutions from Sec.\:III, which serve as the initial conditions for our simulations, we now turn to numerical techniques for evolving the time-dependent Schrödinger-Newton system under accreting black hole potentials. Analytical solutions are intractable due to the integral nature of the self-potential and the exponential mass growth term, necessitating stable and efficient computational methods. We explore two approaches: one directly evolving the wave function $\Psi(r, t)$, which offers physical intuition but can introduce numerical instabilities near the origin; and a more robust alternative using the transformed radial function $\chi(r, t) \equiv r \Psi(r, t)$, which regularizes singularities and preserves norm conservation. The latter proves particularly effective for long-term evolutions, as demonstrated in our results.

Using the transformation $r\Psi(r,t)=\chi(r,t)$ described above, the 1+1 dimensional integro-differential equation governing the spherically symmetric solutions of the scalar field becomes
\begin{align}\label{IntegralEq}
    & i \hbar \frac{\partial}{\partial t} \chi(r,t) 
    = - \frac{\hbar^2}{2m} \frac{\partial^2}{\partial r^2} \chi(r,t) - G m \left( \frac{4 \pi}{r} \int_0^r |\chi(r',t)|^2 \, dr' \nonumber \right.\\
    &\left.+ 4 \pi \int_r^\infty \frac{|\chi(r',t)|^2}{r'} \, dr' 
    + \frac{M_0 e^{\alpha t}}{r} \right) \chi(r,t).
\end{align}
To enhance numerical stability and avoid the singularity at $r=0$ arising from the $\frac{1}{r}$ terms in the potential due to the black hole mass $M_0$, we introduce the transformation $\chi(r, t) \equiv r \Psi(r, t)$. This substitution recasts the original equation into a form with a standard one-dimensional Laplacian $\frac{\partial^2}{\partial r^2}$, while regularizing the behavior near the origin (see Eq.~\eqref{IntegralEq}).
An additional advantage is that $|\chi(r, t)|^2$ itself represents the radial probability density up to a normalization constant, owing to the unitary nature of the evolution. Specifically, the total probability is preserved as
\begin{align}
\int_{0}^{\infty} r^2 |\Psi(r, t)|^2 , dr = \int_{0}^{\infty} |\chi(r, t)|^2  dr = \frac{1}{4 \pi},
\end{align}
where the factor of $\frac{1}{4\pi}$ accounts for the spherical symmetry. This formulation makes it straightforward to monitor norm conservation by verifying that $\int_{0}^{\infty} |\chi(r, t)|^2 , dr$ remains constant throughout the simulation, providing a useful diagnostic for numerical accuracy. Furthermore, since the $-\frac{1}{r}$ potential is attractive, we expect the long-time behavior to feature a peak in $|\chi(r, t)|^2$ shifting toward the origin, indicative of a reduction in the size of the soliton core and an increase in its central density, i.e., a squeezing effect on the density profile. We refer to $|\chi(r, t)|^2$ as the \textit{radial density}. In what follows, we detail the numerical method for evolving this system.

We now use the Crank-Nicolson scheme to solve the above 1+1 dimensional differential integral equation. To do so, we define the spatial and temporal grid sizes as $\Delta r$ and $\Delta t$ respectively, and use the index notation $\Psi_k^n(r,t)\equiv \Psi(k\Delta r,n\Delta t)$. Then, we can write Eq.\:\eqref{1plus1} in the Cayley form as \cite{salzman2005investigation, franklin2016dynamics}
\begin{equation}
    \exp\left(\frac{i\Delta t\hat{H}}{2\hbar}\right)\Psi_k^{n+1}=\exp\left(-\frac{i\Delta t\hat{H}}{2\hbar}\right)\Psi_k^{n}.
\end{equation}
After linearization, it can be written as
\begin{equation}
    \Psi_k^{n+1}=(\hat{Q}^{-1}-1)\Psi_k^n,\:\hat{Q}\equiv\frac{1}{2}\left(1+\frac{i\Delta t}{2\hbar}\hat{H}\right).
\end{equation}
Or equivalently, a system of linear equations
\begin{equation}
    \Psi^{n+1}=\chi^n-\Psi^n,\:\chi^n\equiv \hat{Q}^{-1}\Psi^n.
\end{equation}
Here, the radial part of the Laplacian has the form
\begin{subequations}
\begin{align}
\nabla^2&\equiv \frac{\partial^2}{\partial r^2}+\frac{2}{r}\frac{\partial}{\partial r}\:\:\mbox{for}\:\:r>0,\\
\nabla^2&=3\frac{\partial^2}{\partial r^2}\:\:\:\:\:\:\:\:\:\:\:\:\:\mbox{for}\:\:r=0,
\end{align}
\end{subequations}
where the finite difference discretization scheme for the radial part of the Laplacian is (see Appendix\:\ref{Boundary})
\begin{subequations}
\begin{align}
    \nabla^2 \Psi_k^n&=\frac{1}{(\Delta r)^2}\left(\frac{k+1}{k}\Psi_{k+1}^n-2\Psi_{k}^n+\frac{k-1}{k}\Psi_{k-1}^n\right)\:\:\mbox{for}\:\:k>0,\\
    \nabla^2 \Psi_k^n&=\frac{6}{(\Delta r)^2}\left(\Psi_1^n-\Psi_0^n\right)\:\:\:\:\:\:\:\:\:\:\:\:\:\:\:\:\:\:\:\:\:\:\:\:\:\:\:\:\:\:\:\:\:\:\:\:\:\:\:\:\:\:\mbox{for}\:\:k=0.
\end{align}
\end{subequations}
Besides, the discretization of the self-potential and the black hole potential have the form
\begin{subequations}
\begin{align}
    \Phi_k^n&\approx -4\pi G(\Delta r)^2\left(\frac{1}{k}\sum_{i=0}^{k-1}|\Psi_i^n|^2i^2+\sum_{i=k}^{N-1}|\Psi_i^n|^2i\right)\equiv -4\pi G(\Delta r)^2f_k^n,\\
    V_k^n&= -GM_0\frac{\exp(\alpha n\Delta t)}{k(\Delta r)}\equiv -GM_0g_k^n.
\end{align}
\end{subequations}
Away from the origin and the point before spatial infinity of the spatial grid, i.e., $k\neq 0,N-1$, one obtains the finite difference equations
\begin{subequations}
\begin{gather}
    \hat{Q}\chi_k^n\equiv a_k\chi_{k-1}^n+b_k\chi_k^n+c_k\chi_{k+1}^n,\\
    a_k\equiv \beta\frac{k-1}{k},\:b_k\equiv \frac{1}{2}-2\beta+\gamma f_k^n+\delta g_k^n,\:c_k\equiv \beta\frac{k+1}{k},\label{Coefficients1}\\
    \beta\equiv -\frac{i\hbar}{8m}\frac{\Delta  t}{(\Delta r)^2},
    \gamma\equiv \frac{-i\pi Gm(\Delta r)^2\Delta t}{\hbar},
    \delta=\frac{-iGM_0m\Delta t}{4\hbar}\label{Coefficients2}.
\end{gather}
\end{subequations}
Similarly, at the point directly before spatial infinity of the spatial grid, i.e., $k=N-1$, one obtains the finite difference equation
$\hat{Q}\chi_{N-1}^n=a_{N-1}\chi_{N-2}^n+b_{N-1}\chi_{N-1}^n$, where the coefficients $a_{N-1}$ and $b_{N-1}$ are given by Eqs.\:\eqref{Coefficients1} - \eqref{Coefficients2} by substituting $k$ for $N-1$. Finally, at the origin, i.e., $k=0$, one obtains the finite difference equation $\hat{Q}\chi_0^n\equiv b_0\chi_0^n+c_0\chi_0^n$, where $b_0\equiv\frac{1}{2}-6\beta+\gamma f_0^n+\delta g_0^n$, and $c_0\equiv 6\beta$.
Denoting $\tau\equiv t/t_E$, one obtains the dimensionless finite difference equation at the $(n+1)$th time step, $\tilde{\Psi}_k^{n+1}=(\hat{Q}^{-1}-1)\tilde{\Psi}_k^n$, where
\begin{subequations}
\begin{gather}
    \hat{Q}\tilde{\Psi}_k^n=\tilde{a}_k\tilde{\Psi}_{k-1}^n+\tilde{b}_k\tilde{\Psi}_k^n+\tilde{c}_k\tilde{\Psi}_{k+1}^n,\tilde{\Psi}_k^n\equiv \sqrt{4\pi G}t_E\Psi_k^n,\\
    \tilde{a}_k\equiv \tilde{\beta}\frac{k-1}{k},\tilde{b}_k\equiv\frac{1}{2}-2\tilde{\beta}+\tilde{\gamma}\tilde{f}_k^n+\delta\tilde{g}_k^n,\tilde{c}_k\equiv\tilde{\beta}\frac{k+1}{k},\\
    \tilde{\alpha}\equiv\alpha t_E,\tilde{\beta}=\frac{-i\Delta\tau}{8(\Delta u)^2},\tilde{\gamma}\equiv \frac{-i(\Delta u)^2\Delta\tau}{4},\tilde{\delta}\equiv \frac{-i\lambda\Delta\tau}{4}.
\end{gather}
\end{subequations}
Hence, $\hat{Q}$ can be represented as a sparse $N$ by $N$ matrix in a tridiagonal form
\begin{equation}
    \hat{Q}\equiv 
    \begin{pmatrix}
b_0 & c_0 & 0 & 0 & \cdots & 0\\
a_1 & b_1 & c_1 & 0 & \cdots & 0\\
0 & a_2 & b_2 & c_2 & \cdots & 0\\
\vdots  & \vdots  & \vdots & \ddots  & \vdots & \vdots\\
0 & 0 & \cdots & a_{N-2} & b_{N-2} & c_{N-2}\\
0 & 0 & \cdots & 0 & a_{N-1} & b_{N-1}
\end{pmatrix}.
\end{equation}
Notice that only the coefficients $b_k$ are time-dependent, since they contain a sum over the self-potential and the time-dependent black hole potential. Here, the initial condition for the scalar field $\Psi(r,t)$ is fixed by setting $\Psi(r,0)=\phi(r)$, where $\phi(r)$ is the steady-state spherical solution for $\alpha=0$, which corresponds to a time independent black hole potential with a mass $M(t)\equiv M_0$.

We now provide the second method, based on the transformation $\chi(r,t) \equiv r\Psi(r,t)$, which is often numerically more stable near the origin. This approach avoids the explicit $1/r$ terms in the Laplacian for $\Psi$ by recasting the equation in a form resembling a one-dimensional quantum problem, while naturally enforcing $\chi(0,t)=0$ to maintain finite $\Psi$ at the center. Substituting into Eq.~\eqref{1plus1}, the evolution equation becomes
\begin{align}\label{chi-evolution}
    & i \hbar \frac{\partial}{\partial t} \chi(r,t) 
    = - \frac{\hbar^2}{2m} \frac{\partial^2}{\partial r^2} \chi(r,t) - G m \left( \frac{4 \pi}{r} \int_0^r |\chi(r',t)|^2 \, dr' \nonumber \right.\\
    &\left.+ 4 \pi \int_r^\infty \frac{|\chi(r',t)|^2}{r'} \, dr' 
    + \frac{M_0 e^{\alpha t}}{r} \right) \chi(r,t).
\end{align}
where the Laplacian is now in its reduced 1D radial form, and the wavefunction is evolved directly in terms of $\chi(r,t)$. Note that the first integral simplifies due to cancellation of $r'$ factors, reflecting the spherical volume element.
To solve this numerically, we discretize the spatial domain from $r=0$ to a large finite radius $r_{\max} = (N-1)\Delta r$, chosen such that boundary effects are negligible, and time from $t=0$ to the desired evolution time, using uniform grid spacings $\Delta r$ and $\Delta t$. We denote the discrete values as $\chi_k^n = \chi(k\Delta r, n\Delta t)$, with $k=0,1,\dots,N-1$ and $n=0,1,\dots$. The initial condition is set as $\chi_k^0 = k\Delta r \cdot \phi(k\Delta r)$, where $\phi(r)$ is the steady-state ground-state solution from Section III corresponding to $\alpha=0$.

The Crank–Nicolson scheme, which is unconditionally stable and second-order accurate in both space and time, is applied to preserve unitarity and energy conservation in the evolution:
\begin{equation}
    \left( I + \frac{i\Delta t}{2\hbar} \hat{H}^n \right)\chi^{n+1} = \left( I - \frac{i\Delta t}{2\hbar} \hat{H}^n \right)\chi^{n},
\end{equation}
where $\hat{H}^n$ is the time-dependent Hamiltonian evaluated at time step $n$, acting on the vector $\chi^n = (\chi_0^n, \chi_1^n, \dots, \chi_{N-1}^n)^T$ as
$$    \hat{H}^n = -\frac{\hbar^2}{2m} D^{(2)} + \hat{V}^n.$$
Here, $D^{(2)}$ is the second-order central finite difference matrix approximating the second derivative, with elements for interior points $k=1,\dots,N-2$
$$    (D^{(2)}\chi)_k = \frac{\chi_{k+1} - 2\chi_k + \chi_{k-1}}{(\Delta r)^2}.$$
At the boundaries, we impose $\chi_0^n = 0$ to enforce regularity at the origin, and $\chi_{N-1}^n \approx 0$ approximating decay at infinity, with $r_{\max}$ sufficiently large. For the origin ($k=0$), the stencil is adjusted using a forward difference or ghost point to maintain second-order accuracy, but since $\chi_0^n=0$, it simplifies the system.
The potential $\hat{V}^n$ is a diagonal matrix with entries $V_k^n$ approximating the integral terms via trapezoidal rule for accuracy
$$    V_k^n = -Gm \left[ \frac{4\pi}{r_k} \sum_{j=0}^{k} |\chi_j^n|^2 \Delta r + 4\pi \sum_{j=k}^{N-1} \frac{|\chi_j^n|^2}{r_j} \Delta r + \frac{M_0 e^{\alpha n\Delta t}}{r_k} \right],$$
where $r_k = k \Delta r$. Notice that $r_0=0$ is handled by limiting behavior or skipping in sums, as the $1/r_k$ term at $k=0$ is regularized by the $\chi$ formulation. The sums approximate the integrals with $\mathcal{O}(\Delta r^2)$ error.

This formulation yields a tridiagonal linear system at each time step
$$    A\chi^{n+1} = B\chi^n,$$
where $A = I + \frac{i\Delta t}{2\hbar} \hat{H}^n$ and $B = I - \frac{i\Delta t}{2\hbar} \hat{H}^n$ are sparse tridiagonal matrices arising from the second-derivative stencil and diagonal potential. The system is solved efficiently using the Thomas algorithm, which is a specialized Gaussian elimination for tridiagonals, with computational cost $\mathcal{O}(N)$ per step.

The $\chi(r,t)$ approach is advantageous because it regularizes the $1/r$ singularity near the origin inherent in the $\Psi(r,t)$ formulation, while preserving the norm through
$$    \int_0^\infty |\chi(r,t)|^2\,dr = \int_0^\infty |\Psi(r,t)|^2 r^2\,dr = \frac{1}{4\pi},$$
where the factor $1/(4\pi)$ accounts for spherical symmetry in the normalization. In all simulations, we explicitly renormalize $\chi^n$ after each step to enforce unitarity, as minor numerical drift can occur due to finite precision. We monitor this norm as a consistency check, ensuring it remains constant within machine epsilon throughout the evolution.
Compared to the $\Psi$-based method, we found this $\chi$-based approach to be more stable, particularly for small $\Delta r$ near the origin and long evolution times, as it avoids amplification of numerical errors from the spherical Laplacian terms. Simulations using this method, with parameters such as $\Delta t = 5 \times 10^{-4}$ in dimensionless units over $2 \times 10^4$ steps, are presented in Fig.~3, demonstrating convergence (see App \ref{sec:con} for tests).

As a result of the improved Crank–Nicolson scheme, Fig.\:3 presents the radial densities $|\tilde{\chi}(u, \tau)|^2 \equiv 4\pi G \hbar^2 / (m^2 c^4) |\chi(r, t)|^2$ as functions of the normalized radial distance $u \equiv m c r / \hbar$ and dimensionless time $\tau \equiv m c^2 t / \hbar$, corresponding to an initial 0-node soliton solution within the Schrödinger-Newton framework. Results are presented for dimensionless black hole masses $\lambda \equiv G M_0 m / (\hbar c) = 0, 0.1, 0.5,$ and $1.0$ respectively, with a fixed accretion rate $\alpha = 0.1$. The initial profile at $\tau=0$ corresponds to the steady-state equilibrium solution obtained via the shooting method, satisfying the asymptotic boundary condition $\tilde{\phi}(\infty) = 0$. 

As the supermassive black hole accretes according to $M(t) = M_0 e^{\alpha t}$, the density profile exhibits a pronounced squeezing effect: the inner head region ($u \lesssim 2$) undergoes rapid compression, increasing the central density and reducing the core radius, while the outer tail region ($u \gtrsim 3$) elongates and develops oscillatory features at later times ($\tau = 5, 10$). These oscillations arise from wave-like interference patterns in the ultralight scalar field, driven by the interplay between the time-dependent attractive potential $-G M(t)/r$ and the repulsive quantum pressure term $-\hbar^2 \nabla^2 / (2m)$. Norm conservation is preserved, with mass redistribution manifesting as damped density waves propagating outward, extending the influence to larger halo scales. The effects are amplified for higher $\lambda$, indicating stronger perturbations for more massive initial black holes. Numerical evolution employs the Crank-Nicolson scheme with $\Delta t = 5 \times 10^{-4}$ over $2 \times 10^4$ steps; convergence is verified in App.\:C, confirming the oscillations as physical rather than numerical artifacts.

To further elucidate the role of the accretion rate in modulating the dynamical response, Fig.\:4 depicts the normalized radial densities $|\tilde{\chi}(u, \tau)|^2 \equiv 4\pi G \hbar^2 / (m^2 c^4) |\chi(r, t)|^2$ at the final normalized time $\tau = 10$, evolved from the initial 0-node configuration under the influence of an accreting supermassive black hole. Each subplot corresponds to a fixed dimensionless gravitational coupling $\lambda = GM_0 m / (\hbar c) \in {0, 0.1, 0.5, 1.0}$, with comparative profiles for varying accretion rates $\alpha \in {0, 0.05, 0.1}$ overlaid to highlight their amplifying effects on core squeezing. In the absence of accretion ($\alpha = 0$), the density profile remains largely static, exhibiting minimal deviation from the initial equilibrium due to the constant potential $V(r, t) = -GM_0/r$; however, increasing $\alpha$ accelerates the black hole mass growth $M(t) = M_0 e^{\alpha t}$, intensifying the attractive perturbation and yielding sharper central peaks, contracted core radii, and persistent oscillatory tails, particularly pronounced at higher $\lambda$ where initial gravitational influences are stronger. These simulations, conducted via the Crank-Nicolson method with $\Delta t = 5 \times 10^{-4}$ over $2 \times 10^4$ steps and validated for convergence in App.\ref{sec:con}, emphasize the sensitivity of fuzzy dark matter structures to black hole accretion dynamics, providing insights into potential resolutions of the cusp-core discrepancy in galactic observations.

\section{Restrictions on fuzzy dark matter particle mass}
\subsection{The mass-radius relation for the soliton/ boson star}
For an intuiting understanding of nonlinear, bound objects like solitons or boson stars, one can start with the analysis of different energy scales. In particular, solitons are those compact objects in which quantum pressure balances gravitational potential energy. i.e.,
\begin{equation}\label{MassRadiusRelation}
\frac{GM_s}{R_s}\propto \frac{1}{2m^2}\frac{\nabla^2\sqrt{\rho}}{\sqrt{\rho}}\propto \frac{1}{m^2R_s^2},
\end{equation}
where $\rho\equiv m|\phi|^2$ is the mass density, $R_s$ is the radius of the soliton, and $M_s$ is the total mass of the soliton given by \cite{davies2020fuzzy}
\begin{equation}\label{Totalmass}
M_s\equiv \int_0^{\infty} |\phi|^2 4\pi r^2dr = \frac{\hbar c}{Gm}\int_0^{\infty} \tilde{\phi}^2u^2du \equiv \frac{\hbar c\tilde{M}_s}{Gm},
\end{equation}
where $u\equiv r/L_0$, $L_0\equiv \hbar/(mc)$, $\tilde{\phi}\equiv \sqrt{4\pi G}L_0\phi/c$, and $\tilde{M}_s$ is the normalized total mass of the soliton. The constant $\hbar c/(Gm)$ sets the mass of the soliton and is given by
\begin{equation}
\frac{\hbar c}{Gm} \approx 1.336\times 10^{12}M_\odot\left(\frac{10^{-22}\mbox{eV}}{m}\right).
\end{equation}
One may introduce the dimensionless ratio of the black hole mass to the soliton mass, $\eta\equiv \lambda/\tilde{M}_s=M_0/M_s$, where $M_0$ is the mass of the supermassive black hole, and $\lambda$ sets the dimensionless mass of the black hole. Clearly, the mass ratio $\eta$ is invariant under scaling, and is the main factor which determines the density profile of the soliton. 

From Eq.\:\eqref{MassRadiusRelation}, one deduces that the total mass of a soliton is inversely proportional to its core size, i.e., \cite{hui2021wave} 
\begin{equation}
M_s R_s\propto \frac{1}{Gm^2}\propto 2.2\times 100\mbox{pc}\times 10^9M_\odot\left(\frac{10^{-22}\mbox{eV}}{m}\right)^2.
\end{equation}
Using the data $1\mbox{pc} = 3.085677581\times 10^{16}\mbox{m}$, one obtains the mass-radius relation for a soliton in the dimensionless unit
\begin{figure}[tbp]
\begin{center}  \includegraphics[width=1\columnwidth]{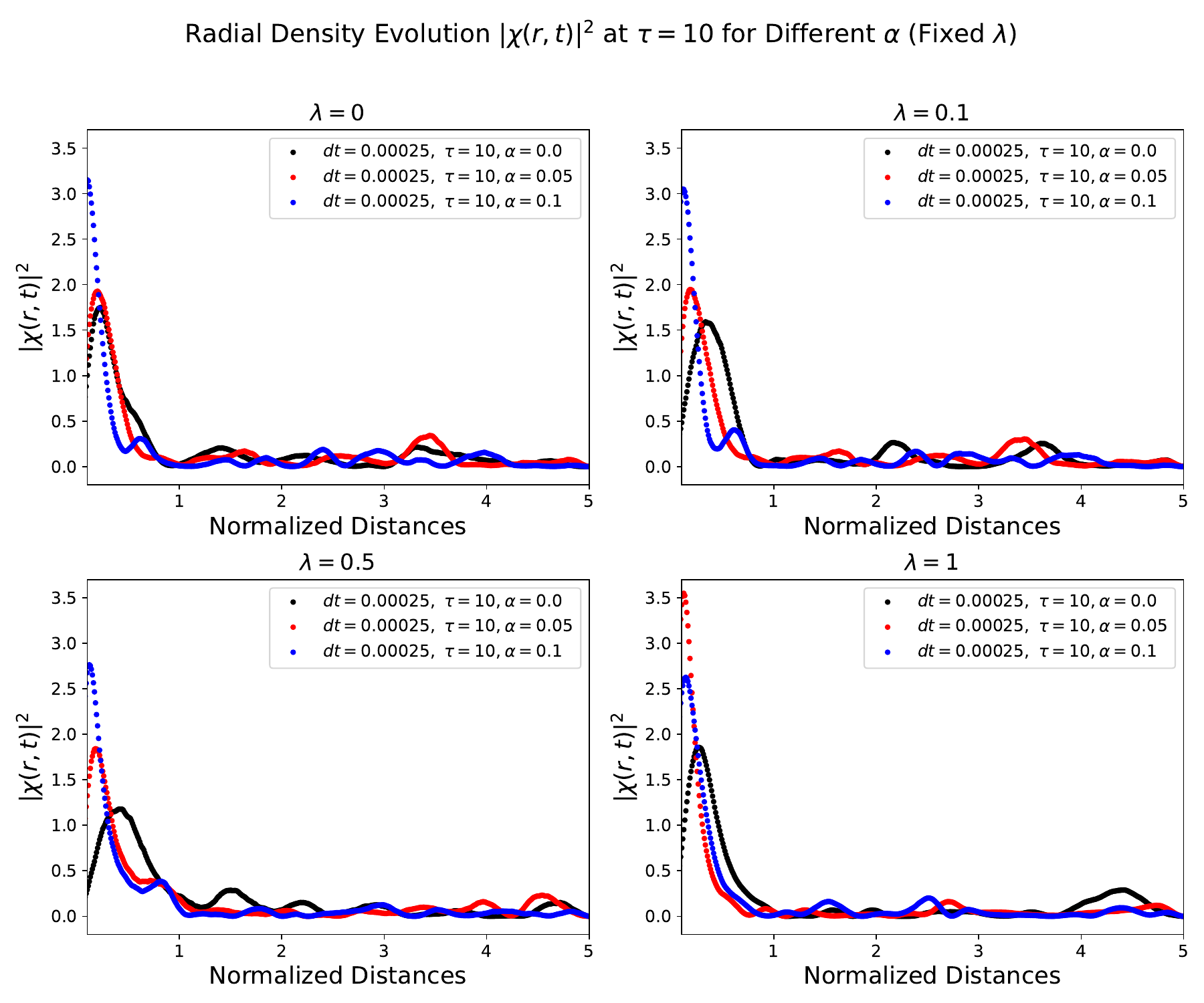}%
\caption{Normalized radial densities $|\tilde{\chi}(u,\tau)|^2 \equiv 4\pi G\hbar^2/(m^2c^4) |\chi(r,t)|^2$ at final time $\tau = 10$, evolved from an initial 0-node (ground state) configuration. Each subplot corresponds to a fixed value of the gravitational coupling $\lambda = 0$, $0.1$, $0.5$, and $1$. Within each subplot, results for different self-interaction strengths $\alpha = 0$, $0.05$, and $0.1$ are shown for comparison.}
\label{fig:Fig4}
\end{center}
\end{figure}

\begin{equation}
\tilde{M}_sU_s = 2.57436836743.
\end{equation}

\subsection{Soliton condensation: the relationship between the soliton mass and the halo mass}\label{SolitonConden}
On the large scale of the Universe, it was first pointed out by Schive et al. \cite{schive2014cosmic, schive2014understanding} from the cosmological simulations that, virialized halos tend to have a core with a soliton mass that scales with the halo mass as:
\begin{equation}
M_s\propto a^{-1/2}M_{h}^{1/3},
\end{equation}
where $a$ is the cosmic scale factor, $M_s$ is the soliton mass, and $M_h$ is the dark matter halo mass. Employing the scaling relation at the present age of the Universe ($a=1$), one obtains
\begin{equation}
M_s \approx 1.25\times 10^9M_\odot\left(\frac{M_h}{10^{12}M_\odot}\right)^{1/3}\left(\frac{10^{-22}\mbox{eV}}{m}\right).
\end{equation}
As a remark, a dark matter halo is said to be virialized when it has reached an equilibrium between the kinetic and potential energy, and and obeys the virial theorem \cite{hui2021wave}. The virial radius of a dark matter halo is the radius that which the density of the halo equals $200$ times the critical density of the Universe. The mass contained inside the virial radius is regarded as the total mass of the dark matter halo.

\subsection{The relationship between the soliton mass and the FDM particle mass}
The restriction of the range of dark matter soliton mass mainly comes from the cosmological setting. In order that the soliton does not collapse directly to a black hole, there is an upper limit of the soliton mass \cite{hui2021wave}: $GM/R\lessapprox 1$. Substitution in the mass-radius relation for the soliton yields the maximum soliton mass $M_2$ \cite{hui2017ultralight}
\begin{figure}[tbp]
\includegraphics[width=0.75\columnwidth]{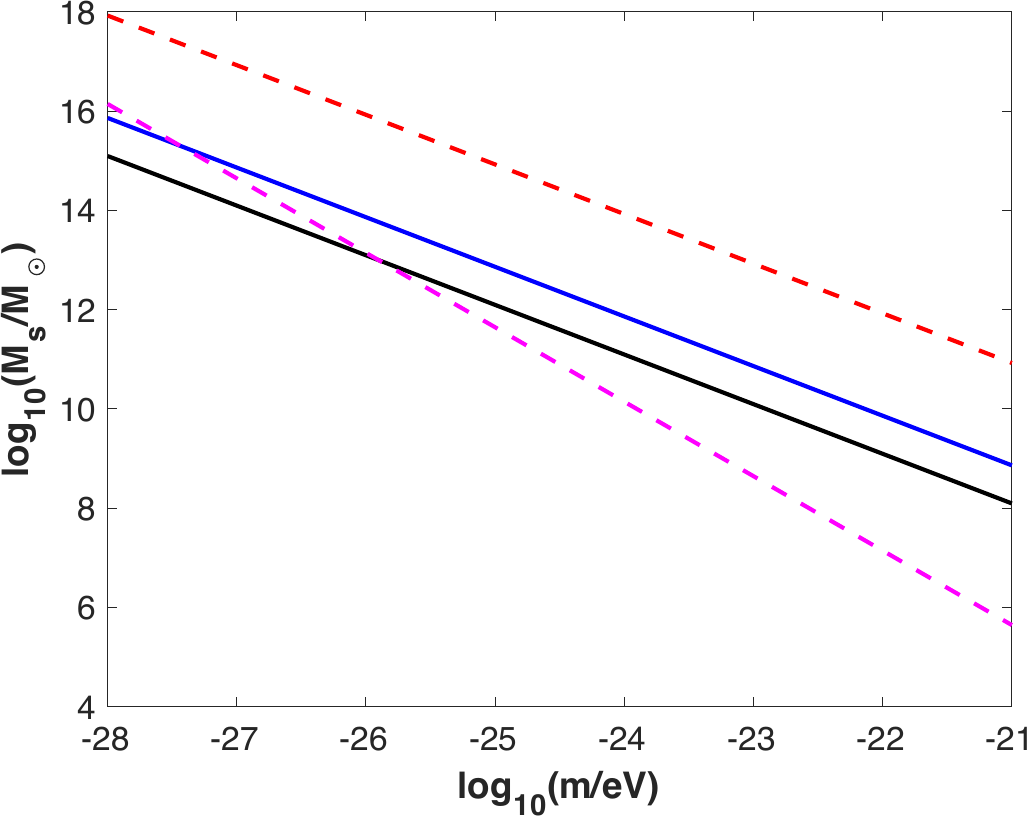}
\caption{The log-log plot of the soliton mass $M_s$ against the fuzzy dark matter mass $m$. The total mass of the soliton at the center of the Milky Way and M87 are shown with dark and blue solid curves respectively. The soliton mass is obtained from the relationship between the soliton mass and the mass of halos where the galaxies reside. The upper and the lower bounds of the dark matter soliton mass are shown by red and magenta broken lines respectively.}
\label{Fig6a}
\end{figure}\begin{equation}
M_2\propto \frac{1}{Gm}\propto 8.46\times 10^{11}M_\odot\left(\frac{m}{10^{-22}\mbox{eV}}\right).
\end{equation}
Similarly, there is a lower limit of the soliton mass. The soliton ceases to exist if the central density is lower than $200$ times the critical density of the Universe, as we have remarked above. Numerically, the minimum soliton mass $M_1$ is \cite{hui2017ultralight}
\begin{equation}
M_1\propto 1.40\times 10^7M_\odot\left(\frac{m}{10^{-22}\mbox{eV}}\right)^{-3/2}.
\end{equation}
Finally, using the following data
\begin{subequations}
\begin{align}
\frac{GM_\odot}{c^2} &= 14.7662563825\times 10^8 \mu\mbox{m},\\
 \hbar c &= 0.1973269804\mbox{eV}\times\mu\mbox{m},
\end{align}
\end{subequations}
one readily obtains from Eq.\:\eqref{Totalmass} the numerical relationship between the soliton mass and the FDM particle mass
\begin{equation}
M_s/M_\odot = 1.336337765\times 10^{-x-10}\times\tilde{M}_s,
\end{equation}
where $x\equiv \log_{10}(m/\mbox{eV})$, and $\tilde{M}_s$ is the normalized total mass of the soliton. 

\begin{figure}[tbp]
\includegraphics[width=0.75\columnwidth]{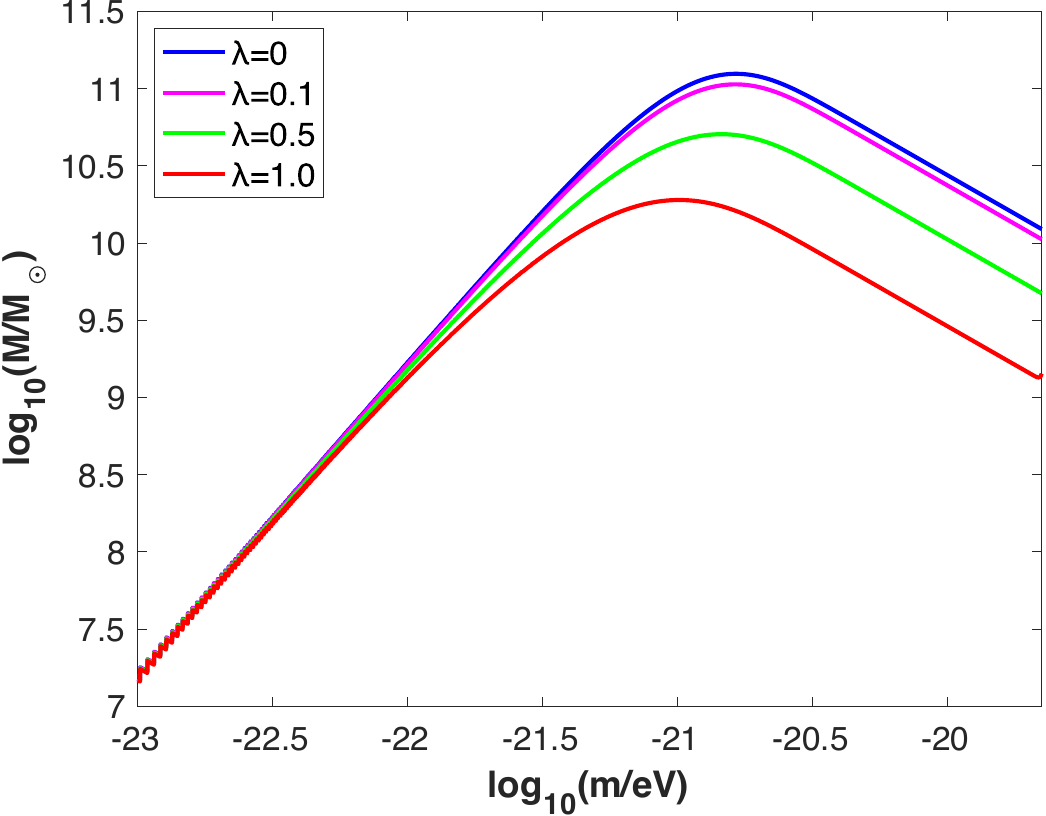}
\caption{The log-log plot of the enclosed soliton mass $M$ within a radius of $r\approx 0.01\mbox{pc}$ near the central supermassive blackhole against the fuzzy dark matter mass $m$. Here, the dimensionless mass of the supermassive blackhole is chosen to be $0$, $0.1$, $0.5$ and $1.0$ respectively.}
\label{Fig6b}
\end{figure}

As illustrated in Fig.\:\ref{Fig6a}, for fuzzy dark matter particles with masses $m \lesssim 10^{-26} \, \mathrm{eV}$, the predicted total mass of the FDM soliton at the Galactic center falls below the established lower bound $M_1$, which arises from the cosmological requirement that the central soliton density must exceed approximately 200 times the critical density of the Universe to maintain structural stability \cite{hui2017ultralight}. This discrepancy implies that exceedingly ultralight bosons would yield diffuse, unstable cores incompatible with the observed properties of Milky Way-like halos, potentially erasing substructure on scales inconsistent with virialized galactic systems. Consequently, to ensure consistency with empirical halo mass relations and cosmological simulations, the FDM particle mass must satisfy $m \gtrsim 10^{-26} \, \mathrm{eV}$. Nevertheless, complementary probes—such as Lyman-$\alpha$ forest observations, ultra-faint dwarf galaxy demographics, and stellar heating in the Milky Way disc—impose more stringent lower limits, typically ranging from $10^{-22} \, \mathrm{eV}$ to $10^{-21} \, \mathrm{eV}$ \cite{irvsic2017first, church2019heating, nadler2021dark, ferreira2021ultra}, thereby narrowing the viable parameter space and underscoring the importance of multi-messenger constraints in validating wave-like dark matter paradigms.

As illustrated in Fig.\:\ref{Fig6b}, we compare the enclosed mass profiles $M(<r)$ of fuzzy dark matter (FDM) soliton cores in proximity to central supermassive black holes (SMBHs) with masses spanning $M_0 \sim 10^5$--$10^{10} \, M_\odot$, corresponding to dimensionless coupling parameters $\lambda \equiv G M_0 m / (\hbar c)$ ranging from 0 to 1, where $m$ denotes the ultralight boson mass. These profiles reveal the gravitational influence of the SMBH on the soliton structure, with $\lambda$ encapsulating the interplay between the black hole's point-mass potential and the wave-like quantum pressure inherent to FDM. For boson masses $m \lesssim 10^{-22.5} \, \mathrm{eV}$, the mass profiles remain largely unaffected by variations in $m$, as the corresponding $\lambda$ values are sufficiently small ($\lambda \ll 1$) for typical galactic SMBH masses, rendering the black hole's perturbation negligible compared to the soliton's self-gravity. In contrast, for $m \gtrsim 10^{-22.5} \, \mathrm{eV}$, the profiles exhibit significant modulation with $m$, as larger boson masses yield proportionally larger $\lambda$, enhancing the SMBH's compressive effect on the core and resulting in steeper density gradients and reduced core radii. This threshold of $m \approx 10^{-22.5} \, \mathrm{eV}$ marks the regime where the de Broglie wavelength $\lambda_\mathrm{dB} \sim \hbar / (m v)$ becomes sufficiently compact (with virial velocity $v \sim 10^{-3} c$) for the SMBH to dominate the central dynamics, consistent with numerical solutions of the Schrödinger-Newton equations that show profile compression scaling with $\lambda$.

Surrounding an SMBH, the enclosed soliton mass at fixed radii can be suppressed by up to two orders of magnitude when the black hole mass approaches or exceeds the intrinsic soliton mass ($M_0 \sim M_s$), as the enhanced gravitational binding disrupts the equilibrium configuration, redistributing FDM density inward and potentially alleviating observational tensions in high-density galactic centers, such as those probed by stellar kinematics or gas dynamics. This squeezing phenomenon not only refines predictions for FDM viability but also offers testable signatures in multi-wavelength observations of galactic nuclei, including those from the Event Horizon Telescope and future gravitational wave detectors.

\section{Conclusion}
In this study, we numerically solved the Schrödinger-Newton equations to investigate the density profiles of fuzzy dark matter (FDM) soliton cores in galactic halos perturbed by an accreting supermassive black hole (SMBH), modeled as a time-dependent point mass $M(t) = M_0 e^{\alpha t}$. Our results demonstrate a pronounced squeezing effect on the soliton structure, manifesting as a reduction in core radius and an enhancement in central density. This effect intensifies with higher initial black hole masses (parameterized by the dimensionless $\lambda \equiv G M_0 m / (\hbar c)$) and accretion rates ($\alpha$), as illustrated in Figs. 3--6. Furthermore, extended time evolutions reveal oscillatory behavior in the outer tail regions at later times ($\tau \approx 10$), characterized by alternating density peaks and troughs extending to normalized distances $u \gtrsim 5$. These oscillations, verified as physical through convergence tests (App.\:C), arise from wave-like interference in the ultralight scalar field $\Psi(r, t)$, driven by the interplay between the growing attractive potential $-G M(t)/r$ and repulsive quantum pressure from the kinetic term $-\hbar^2 \nabla^2 / (2m)$.

Table \ref{tab:table1} enumerates the lowest discrete eigenvalues $\epsilon_n$ of the time-independent Schrödinger-Newton system for varying $\lambda$, underscoring the discrete spectrum enforced by the asymptotic boundary condition $\tilde{\phi}(\infty) = 0$. Unlike linear analogs, achieving exact exponential decay demands high numerical precision; here, we retained 16 digits to ensure accuracy within $10^{-6}$ relative error. The ground-state ($n=0$) profiles, serving as initial conditions, exhibit flat central densities ($\rho \propto r^0$), aligning with FDM's resolution to the cusp-core problem in dwarf galaxies \cite{hui2017ultralight, schive2014cosmic}.

The observed dynamics highlight FDM's quantum coherence, distinguishing it from classical cold dark matter (CDM). Short-term squeezing ($\tau \lesssim 5$) increases central densities to $\rho_c \sim 10^7 - 10^9 M_\odot / \rm{kpc}^3$, potentially bridging observational discrepancies in galactic cores \cite{burkert2015structure, jones2021fuzzy}. At longer timescales, the head-tail dichotomy emerges: adiabatic compression steepens the inner profile toward cusp-like forms, while outward-propagating density waves elongate tails and dissipate energy, preserving norm conservation $\int_0^\infty |\chi(r, t)|^2 dr = 1/(4\pi)$. These waves, propagating at de Broglie velocities $\lambda_{\rm dB} \sim \hbar / (m v) \approx 1$ kpc for $v \sim 10$ km/s, influence halo-scale structures and may link to suppressed substructure formation \cite{hui2017ultralight, ferreira2021ultra}. Echoing superradiant instabilities in boson clouds around rotating black holes \cite{baryakhtar2017black, cardoso2004black}, our non-relativistic, spherically symmetric framework emphasizes Newtonian self-gravitation via Poisson's equation.

In a cosmological context, these findings accentuate FDM's divergence from CDM: whereas CDM halos erode through frictionless infall, FDM's wave mechanics enable long-range perturbations, potentially alleviating small-scale tensions such as the missing satellites problem \cite{bullock2017small} while yielding testable predictions for active galactic nuclei (AGN) hosts. Oscillations could introduce temporal variability in rotation curves or Lyman-$\alpha$ forest constraints, refining FDM mass bounds ($10^{-24} \lesssim m \lesssim 10^{-20}$ eV) \cite{irvsivc2017first}.

Our analysis is limited to the non-relativistic regime, valid at halo scales ($r \gg r_s \sim 10^{-6}$ pc). Future investigations could incorporate fully relativistic treatments of scalar field evolution near accreting SMBHs, including general relativistic effects and self-interactions, to probe damping mechanisms like mode decay or gravitational wave emission. Additionally, self-consistent simulations of the coupled black hole–soliton–halo system would elucidate multi-scale interactions. These extensions hold promise for multi-messenger astronomy, leveraging instruments such as the James Webb Space Telescope and Event Horizon Telescope to probe FDM signatures in galactic centers.

\section{Acknowledgements}
EJK acknowledges financial support from the National Science and Technology Council (NSTC) of Taiwan under Grant No.~NSTC~114-2112-M-A49-036-MY3.

\begin{appendix}
\section{Derivation of the 1+1 dimensional differential integral equation for spherically symmetric solutions}
For spherically symmetric solutions, one may use the Laplace expansion of the inverse distance between the two points $\mathbf{r}$ and $\mathbf{r}^\prime$
\begin{equation}
\frac{1}{|\mathbf{r}-\mathbf{r}^\prime|}=\sum_{l=0}^\infty\frac{4\pi}{2l+1}\frac{r_>^l}{r_<^{l+1}}Y_l^{-m}(\theta,\phi)Y_l^m(\theta^\prime,\phi^\prime),
\end{equation}
where $r_<=\mbox{min}(r, r')$, $r_>=\mbox{max}(r, r')$, and $Y_l^m(\theta,\phi)$ is a normalized spherical harmonic function. Then, an integration of the inverse distance between the two points $\mathbf{r}$ and $\mathbf{r}^\prime$ weighted by the density $|\Psi(r^\prime,t)|^2$ becomes
\begin{align}
   &\sum_{l=0}^\infty\frac{4\pi}{2l+1}\int_0^\infty |\Psi(r^\prime,t)|^2r^{\prime 2}dr^\prime\frac{r_<^l}{r_>^{l+1}}Y_l^{-m}(\theta,\phi)\int Y_l^m(\theta^\prime,\phi^\prime)d\Omega^\prime\nonumber\\
    &=\sum_{l=0}^\infty\frac{4\pi}{2l+1}\int_0^\infty |\Psi(r^\prime,t)|^2r^{\prime 2}dr^\prime\frac{r_<^l}{r_>^{l+1}}Y_l^{-m}(\theta,\phi)\sqrt{4\pi}\delta_{l,0}\delta_{m,0}\nonumber\\
    &=\frac{4\pi}{r}\int_0^r|\Psi(r^\prime,t)|^2r^{\prime 2}dr^\prime + 4\pi\int_r^\infty|\Psi(r^\prime,t)|^2r^\prime dr^\prime,
\end{align}
where $d\Omega^\prime\equiv \sin\theta^\prime d\theta^\prime d\phi^\prime$ is the differential solid angle in the direction $(\theta^\prime,\phi^\prime)$. Using the composite trapezoidal rule, one may discretize the first integral along the radial spatial direction as
\begin{subequations}
\begin{align}
    \frac{4\pi(\Delta r)^2}{k}\left(\sum_{j=1}^{k-1}|\Psi_j(t)|^2j^2+\frac{k^2}{2}|\Psi_k(t)|^2\right)\:\:&\mbox{for}\:\:k>1,\\
    4\pi(\Delta r)^2\cdot\frac{1}{2}|\Psi_1(t)|^2\:\:\:\:\:\:\:\:\:\:\:\:\:\:&\mbox{for}\:\:k=1,
\end{align}
\end{subequations}
where $r\equiv k\Delta r$ and $\Psi_j(t)\equiv\Psi(j\Delta r,t)$. Similarly, one may discretize the second integral along the radial spatial direction as
\begin{align}
    4\pi(\Delta r)^2\left(\frac{k}{2}|\Psi_k(t)|^2+\sum_{j=k+1}^{N-1}|\Psi_j(t)|^2j+\frac{N}{2}|\Psi_N(t)|^2\right)\:\:&\mbox{for}\:\:k<N-1,\\
    4\pi(\Delta r)^2\left(\frac{N-1}{2}|\Psi_{N-1}(t)|^2+\frac{N}{2}|\Psi_N(t)|^2\right)\:\:\:\:\:\:\:\:&\mbox{for}\:\:k=N-1,
\end{align}
where $r_{max}\equiv N\Delta r$ for a large positive integer $N$ represents the spatial infinity. Hence, for $0<k<N-1$, the discretization of the self-potential along the radial spatial direction has the form
\begin{align}
    \Phi_k(t)&=-4\pi G(\Delta r)^2\left(\frac{1}{k}\sum_{j=1}^{k}|\Psi_j(t)|^2j^2\right.\nonumber\\
    &\left.+\sum_{j=k+1}^{N-1}|\Psi_j(t)|^2j+\frac{N}{2}|\Psi_N(t)|^2\right),
\end{align}
where $\Phi_k(t)\equiv\Phi(k\Delta r,t)$ and we used the boundary condition $\Psi_N(t)=0$ at the spatial infinity. In particular, for $k=N-1$, the discretization of the self-potential in the radial direction has the form
\begin{equation}
    \Phi_1(t)=-4\pi G(\Delta r)^2\left(\frac{1}{k}\sum_{j=1}^{N-1}|\Psi_j(t)|^2j+\frac{N}{2}|\Psi_N(t)|^2\right).
\end{equation}
For a non-uniform partition of $[0,r_k]$ such that $0=r_0<r_1<\cdots<r_{k-1}<r_k$, the trapezoidal rule yields the discretization of the first integral along the spatial direction
\begin{equation}
    \frac{4\pi}{r_k}\left[\frac{1}{2}\sum_{j=1}^{k}\left(|\Psi_{j-1}(t)|^2r_{j-1}^2+|\Psi_j(t)|^2r_j^2\right)\Delta r_j\right],
\end{equation}
where $\Delta r_j\equiv r_j-r_{j-1}$. Similarly, for a non-uniform partition of $[r_k,r_{max}]$ such that $r_k<r_{k+1}<\cdots<r_{N-1}<r_N=r_{max}$, the trapezoidal rule yields the discretization of the second integral along the spatial direction
\begin{equation}
    4\pi\left[\frac{1}{2}\sum_{j=k+1}^{N}\left(|\Psi_{j-1}(t)|^2r_{j-1}+|\Psi_j(t)|^2r_j\right)\Delta r_j\right].
\end{equation}
Using the central difference formula, the discretization of the second order derivative of $\Psi(r,t)$ with respect to $r$ at $r_k$ reads
\begin{align}
    &\left.\frac{\partial^2\Psi(r,t)}{\partial r^2}\right|_{r=r_k}=\frac{\Psi^\prime(r_{k+1/2},t)-\Psi^\prime(r_{k-1/2},t)}{r_{k+1/2}-r_{k-1/2}}\nonumber\\
    &=\frac{2}{\Delta r_{k+1}+\Delta r_k}\left(\frac{\Psi(r_{k+1},t)-\Psi(r_k,t)}{\Delta r_{k+1}}-\frac{\Psi(r_k,t)-\Psi(r_{k-1},t)}{\Delta r_k}\right)\nonumber\\
    &=\frac{2\Psi_{k+1}(t)}{(\Delta r_{k+1}+\Delta r_k)\Delta r_{k+1}}-\frac{2\Psi_k(t)}{
    \Delta r_{k+1}\Delta r_k}+\frac{2\Psi_{k-1}(t)}{(\Delta r_{k+1}+\Delta r_k)\Delta r_k}.
\end{align}
Similarly, the discretization of the first order derivative of $\Psi(r,t)$ with respect to r at $r_k$ reads
\begin{equation}
\left.\frac{\partial\Psi(r,t)}{\partial r}\right|_{r=r_k}=\frac{\Psi_{k+1}(t)-\Psi_{k-1}(t)}{\Delta r_{k+1}+\Delta r_k}.
\end{equation}
Hence, the finite difference formula for the radial part of the Laplacian reads
\begin{align}
    &\left.\nabla^2\Psi(r,t)\right|_{r=r_k}\equiv\left. \left(\frac{\partial^2}{\partial r^2}+\frac{2}{r}\frac{\partial }{\partial r}\right)\right|_{r=r_k}=\frac{2\Psi_{k+1}(t)}{(\Delta r_{k+1}+\Delta r_k)\Delta r_{k+1}}\frac{r_{k+1}}{r_k}\nonumber\\
    &-\frac{2\Psi_k(t)}{\Delta r_{k+1}\Delta r_k}+\frac{2\Psi_{k-1}(t)}{(\Delta r_{k+1}+\Delta r_k)\Delta r_k }\frac{r_{k-1}}{r_k}.
\end{align}

\section{Boundary conditions for the scalar field at $r\rightarrow 0$}\label{Boundary}
In this work, the boundary conditions for the scalar field $\Psi(r,t)$ are given by  
\begin{equation}
   \partial_r\Psi(0,t)=0\:\:\mbox{and}\:\:\Psi(\infty,t)=0.
\end{equation}
For the radical part of the Laplacian, one obtains
\begin{equation}
    \lim_{r\rightarrow 0}\nabla^2\Psi(r,t)\equiv \lim_{r\rightarrow 0}\left(\frac{\partial^2}{\partial r^2}+\frac{2}{r}\frac{\partial}{\partial r}\right)\Psi(r,t)=3\frac{\partial^2\Psi}{\partial r^2}(r,t),
\end{equation}
where we have used the L'Hospital's Rule to determine the indeterminate form of $0/0$
\begin{equation}
    \lim_{r\rightarrow 0}\frac{1}{r}\frac{\partial\Psi(r,t)}{\partial r}=\frac{\partial^2\Psi(r,t)}{\partial r^2}.
\end{equation}
The finite difference discretization of the boundary condition at $r\rightarrow 0$ has the form
\begin{equation}
    \lim_{r\rightarrow 0}\frac{\partial\Psi}{\partial r}(r,t)\approx \frac{\Psi_1^n-\Psi_{-1}^n}{2\Delta r}=0,
\end{equation}
which yields the boundary condition $\Psi_1^n=\Psi_{-1}^n$ at arbitrary time steps $n$. Hence, the finite difference discretization of the radical part of the Laplacian has the form
\begin{equation}
    \lim_{r\rightarrow 0}\nabla^2\Psi(r,t)=\frac{3}{(\Delta r)^2}\left(\Psi_1^n-2\Psi_0^n+\Psi_{-1}^n\right)=\frac{6}{(\Delta r)^2}\left(\Psi_1^n-\Psi_0^n\right).
\end{equation}

\section{Convergence of the Crank-Nicolson Method}
\label{sec:con}
\subsection{Zero-node solutions}
\begin{figure}
    \centering
    \includegraphics[width=1\linewidth]{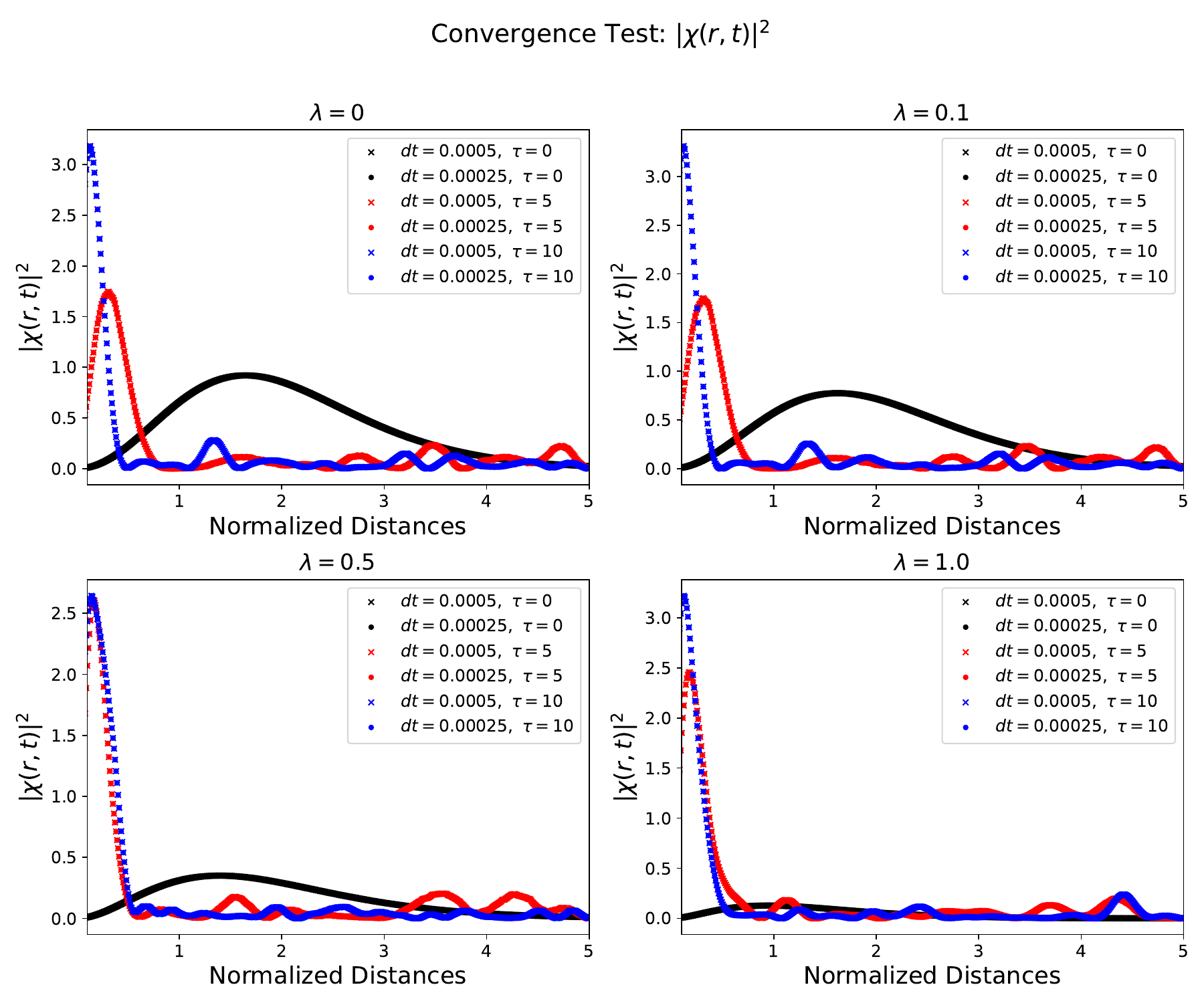}
    \caption{we pick $dt=5\times 10^{-4},$ and evolves for up to  $2\times 10^{4}$ steps and $dt=2.5\times 10^{-4}$ with $4\times 10^{4}$ steps and compared the similuation results. One can see that they fits well. We use 'x' for
$dt=5\times 10^{-4},$ and 'o' for $dt=2.5\times 10^{-4}$ which indicates that our algorithm fits well. Radical densities $|\tilde{\chi}(u,\tau)|^2 \equiv 4\pi G\hbar^2/(m^2c^4)|\chi(r,t)|^2$ as functions of radial distance and time in dimensionless units, for an initial 0-node (ground state) solution. We present results for $\lambda \equiv GM_0m/(\hbar c) = 0$, $0.1$, $0.5$, and $1.0$, with fixed $\alpha = 0.1$ in all cases. }
    \label{fig:convergence}
\end{figure}

To verify the numerical convergence of the Crank-Nicolson scheme employed in our time-dependent simulations, we conduct a timestep refinement analysis for evolutions up to the normalized time $\tau = 10$. In the baseline simulation, we utilize a timestep of $\Delta t = 5 \times 10^{-4}$, requiring $2 \times 10^{4}$ iterations to reach the target time. For the refinement test, we halve the timestep to $\Delta t = 2.5 \times 10^{-4}$, correspondingly doubling the number of iterations to $4 \times 10^{4}$ to maintain the same final time. The resulting radial density profiles $|\tilde{\chi}(u, \tau)|^2$ from both simulations are compared in Figure~\ref{fig:convergence}, where the coarser timestep results are marked with 'x' symbols and the finer timestep results with 'o' symbols. The excellent overlap between the two datasets demonstrates robust convergence, indicating that the chosen baseline timestep is sufficiently small to minimize temporal discretization errors.

This refinement approach is a standard technique for assessing the adequacy of numerical truncation in time-dependent simulations, as it exploits the expected reduction in discretization error with smaller step sizes. For instance, similar timestep convergence tests are routinely applied in quantum many-body simulations to validate Trotterization schemes, where halving the Trotter step size and observing consistent results confirms the reliability of the approximation (see, e.g., Fig.~8 in Ref.~\cite{kuo2024energy}, which illustrates convergence with respect to truncation error in a related operator evolution method).

\begin{figure}
    \centering
    \includegraphics[width=1\linewidth]{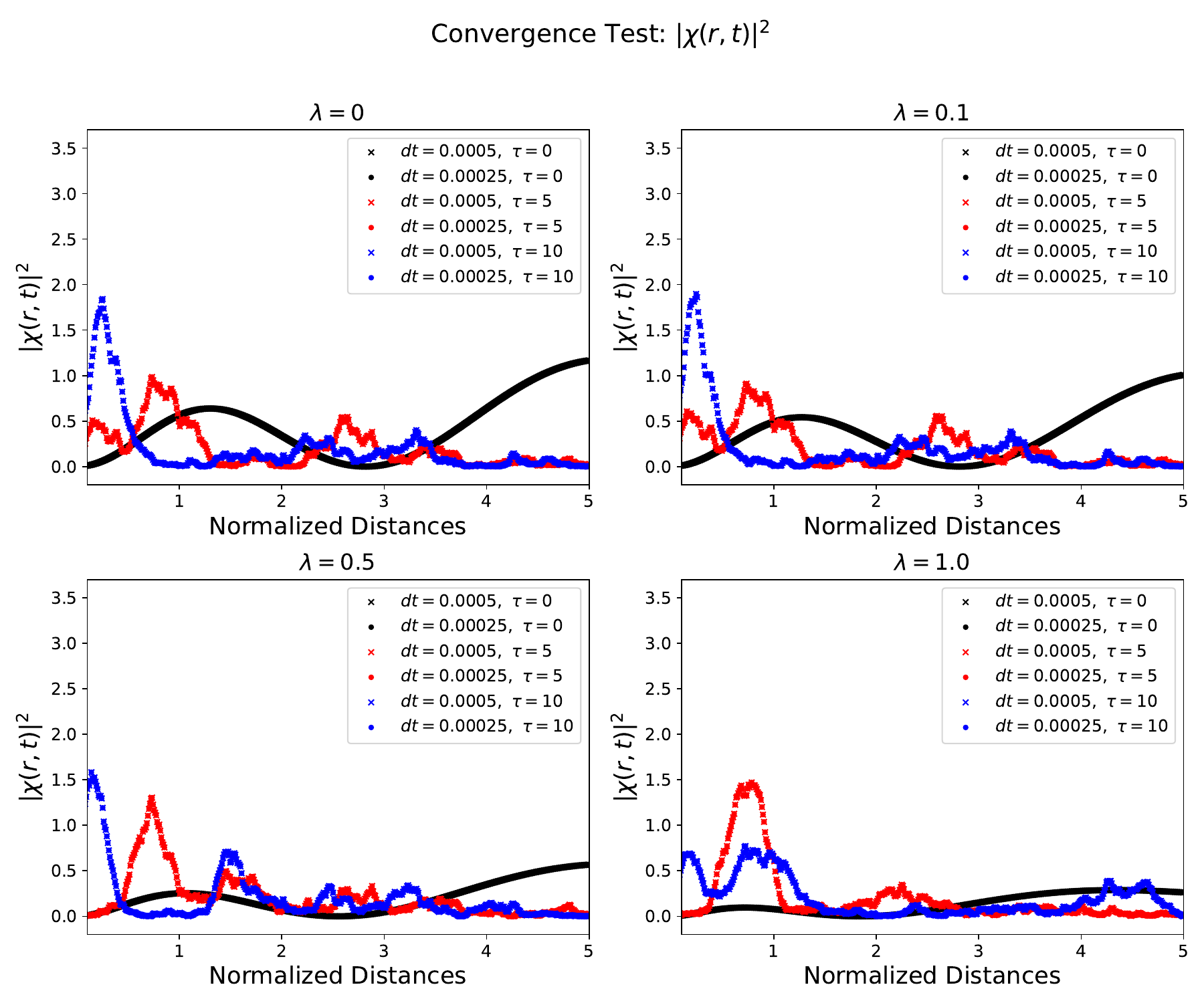}
    \caption{we pick $dt=5\times 10^{-4},$ and evolves for up to  $2\times 10^{4}$ steps and $dt=2.5\times 10^{-4}$ with $4\times 10^{4}$ steps and compared the similuation results. One can see that they fits well. We use 'x' for
$dt=5\times 10^{-4},$ and 'o' for $dt=2.5\times 10^{-4}$ which indicates that our algorithm fits well. Radical densities $|\tilde{\chi}(u,\tau)|^2 \equiv 4\pi G\hbar^2/(m^2c^4)|\chi(r,t)|^2$ as functions of radial distance and time in dimensionless units, for an initial 1-node (ground state) solution. We present results for $\lambda \equiv GM_0m/(\hbar c) = 0$, $0.1$, $0.5$, and $1.0$, with fixed $\alpha = 0.1$ in all cases. }
    \label{fig:1_convergence}
\end{figure}

We extend this convergence test to the first excited state (1-node configuration) at a fixed accretion rate $\alpha = 0.1$, with results presented in Figure~\ref{fig:1_convergence}. Again, the close agreement between the coarse and fine timestep simulations affirms the numerical stability and accuracy of our implementation across different initial states.

\subsection{One-node solutions}
To further substantiate the convergence for the first excited state, we quantify the discrepancies between the coarse and fine timestep runs in Figure 8. The radial densities $|\tilde{\chi}(u, \tau)|^2$ at $\tau=10$ exhibit close alignment across all tested values of $\lambda = 0, 0.1, 0.5,$ and $1.0$, with the coarser timestep ($\Delta t = 5 \times 10^{-4}$, denoted by 'x' symbols) and finer timestep ($\Delta t = 2.5 \times 10^{-4}$, denoted by 'o' symbols) differing by less than $10^{-3}$ in relative error within the core region ($u \lesssim 2$) and remaining below $10^{-2}$ in the oscillatory tails. This level of agreement, achieved after $2 \times 10^4$ and $4 \times 10^4$ steps respectively, underscores the second-order temporal accuracy inherent to the Crank-Nicolson method, even in the presence of the additional radial node that introduces steeper gradients and potential numerical challenges. Moreover, the consistent preservation of the node's position and amplitude without artificial smoothing or phase errors highlights the efficacy of our finite-difference discretization for the nonlocal self-gravitational potential, ensuring reliable long-term evolutions of excited FDM configurations.

\section{Numerical Method for Mass Scaling Relation}

To compute the mass–mass scaling relation of a self-gravitating bosonic system, we numerically solve the coupled Schrödinger–Poisson-like system under spherical symmetry. Specifically, we consider the following ODE system for functions $\psi(t)$ and $\Phi(t)$:

\begin{align}
\psi''(t) &= 2\left( \frac{\Phi(t)}{t} - \epsilon_0 - \frac{\lambda}{t} \right)\psi(t), \\
\Phi''(t) + \frac{2}{t} \Phi'(t) &= \psi(t)^2.
\end{align}
Each $\lambda$ corresponds to a specific eigenvalue $\epsilon_0$ associated with the 0-node configuration, as shown in Table~\ref{tab:table1}. This second-order system is reduced to a system of four first-order equations:
\[
\frac{d}{dt}
\begin{pmatrix}
\psi \\ \psi' \\ \Phi \\ \Phi'
\end{pmatrix}
=
\begin{pmatrix}
\psi' \\
2\left( \frac{\Phi}{t} - \epsilon - \frac{\lambda}{t} \right)\psi \\
\Phi' \\
\frac{1}{t} \psi^2
\end{pmatrix},
\]
with initial conditions $\psi(0) = 1$, $\psi'(0) = 0$, $\Phi(0) = 0$, and $\Phi'(0) = 0$. In numerical implementation, we avoid the singularity at $t=0$ by starting the integration from $t = 10^{-5}$.
We solve this ODE system using MATLAB's \texttt{ode45} solver, which is a Runge–Kutta (4,5) Dormand–Prince method with adaptive step-size control. After obtaining the solution $\psi(t)$, we compute the total mass inside a radius $R$ via:
\[
M(R) = \int_0^R |\psi(t)|^2 t^2 \, dt.
\]
The radius $R$ is parameterized in terms of the particle mass $m$ by
\[
R(m) = 1.56373830646 \times 10^{\log_{10}(m/\mathrm{eV}) + 21}.
\]
For each value of $m$ in the logarithmic range $\log_{10}(m/\mathrm{eV}) \in [-23, -19.65]$, we compute the integrated mass $M(m)$ and define a normalized rescaled quantity:
\[
\log_{10}\left( \frac{M}{M_\odot} \right) = \log_{10}\left( M(m) \times 1.336337765 \times 10^{-\log_{10}(m/\mathrm{eV}) - 10} \right).
\]
This defines the mass–mass scaling curve shown in Fig.\:5. The full numerical procedure follows the MATLAB implementation provided in the supplementary material, where the integral is approximated using the trapezoidal rule \texttt{trapz}, and the ODE is solved once for each pair $(\lambda, \epsilon_0)$. The result is shown in Fig.~\ref{Fig6b}.

\section{Possible Generalizations of the Time-Dependent Potential}
In the main text, we model the growth of the central supermassive black hole using an exponentially increasing mass function,
$$M(t) = M_0 \exp(\alpha t),$$
inspired by the Shakura-Sunyaev thin disk model under Eddington-limited accretion \cite{shakura1973black}. This choice captures the characteristic e-folding timescale associated with radiatively efficient accretion. However, the theoretical framework established in this work, particularly the spherically symmetric Schr\"odinger-Newton equation Eq.~\eqref{IntegralEq},
\begin{align}
i\hbar \partial_t \Psi(r,t) &= -\frac{\hbar^2}{2m r} \partial_r^2 (r \Psi(r,t)) - G m \left( \frac{4\pi}{r} \int_0^r |\Psi(r',t)|^2 r'^2 , dr' \right. \nonumber \\
&\left. + 4\pi \int_r^\infty |\Psi(r',t)|^2 r' , dr' + \frac{M(t)}{r} \right) \Psi(r,t) \nonumber \\
&\equiv \hat{H}(r,t) \Psi(r,t),
\end{align}
accommodates a broader class of time-dependent potentials. Specifically, any sufficiently smooth, positive, and monotonically increasing function $M(t)$ can be incorporated without altering the fundamental structure of the system, provided it remains consistent with the Newtonian approximation at halo scales ($r \gg r_s$).
Such generalizations enable the exploration of diverse accretion histories observed in astrophysical contexts. For instance:
\begin{itemize}
\item A power-law growth model,
$$M(t) = M_0 \left(1 + \beta \frac{t}{t_E}\right)^p,$$
where $p > 0$ and $\beta > 0$ parameterize sub- or super-Eddington phases, potentially arising from variable gas supply or radiative inefficiencies \cite{begelman2002super};

\item A logistic growth model to account for self-regulating feedback mechanisms,
$$M(t) = \frac{M_\infty}{1 + \left( \frac{M_\infty}{M_0} - 1 \right) e^{-\gamma t}},$$
where $M_\infty$ denotes the asymptotic maximum mass and $\gamma$ the growth rate, reflecting saturation due to active galactic nucleus (AGN) outflows or host galaxy evolution \cite{silk1998quasars};

\item A piecewise quenching scenario,
$$M(t) = 
\begin{cases}
M_0 e^{\alpha t} & t \leq t_q, \\
M_0 e^{\alpha t_q} & t > t_q,
\end{cases}$$
where $t_q$ signifies the quenching epoch triggered by feedback processes or environmental factors such as mergers \cite{weinberger2018supermassive}.
\end{itemize}

These alternative forms of $M(t)$ preserve the integro-differential nature of the Schr\"odinger-Newton system but induce distinct perturbations on the fuzzy dark matter soliton core. For example, power-law models may lead to slower initial compression followed by accelerated squeezing, while logistic profiles could stabilize the core density at late times, and quenching scenarios might result in oscillatory transients post-$t_q$. These variations offer opportunities to test fuzzy dark matter against multi-wavelength observations of galactic cores, potentially reconciling tensions with core size and density measurements.

The numerical methodology employed in this study, detailed in Sec.\:IV and App.\:B, is readily extensible to these generalized potentials. The Crank-Nicolson finite-difference scheme provides an unconditionally stable, second-order accurate approach for evolving the wave function under the time-dependent Hamiltonian $\hat{H}(r,t)$. By discretizing the radial domain and approximating the integrals via trapezoidal rules or similar quadrature methods, the scheme transforms each time step into a tridiagonal matrix equation solvable efficiently with the Thomas algorithm. This ensures conservation of the norm and energy to high precision, even for stiff potentials or rapid growth rates, as verified through convergence tests (e.g., Figs. 7 and 8).
Consequently, the framework presented here serves not only as a tool for analyzing exponential accretion but also as a versatile platform for simulating self-gravitating ultralight scalar fields in the presence of arbitrary dynamical sources. Future extensions could incorporate relativistic corrections near the horizon or non-spherical perturbations from galactic bars and mergers, further enhancing the model's applicability to realistic cosmological scenarios.

\end{appendix}

\bibliographystyle{unsrt}
\bibliography{reference_new}

\end{document}